\documentclass[twocolumn, times, tighten]{aastex631}

\usepackage{textcomp}
\usepackage{amsmath}
\usepackage{color}
\usepackage{amssymb}
\usepackage{hyperref}
\usepackage{longtable}
\usepackage{appendix}
\usepackage[T1]{fontenc}
\let\tablenum\relax
\usepackage{siunitx}
\usepackage{xcolor}

\revised{\bf Draft: \today}
\shortauthors{Li et al.}
\shorttitle{A Reverberation mapping study of 6dFGS gJ022550.0-060145}

\begin{document} 

\title{A reverberation mapping study of a highly variable AGN 6dFGS gJ022550.0-060145}

\correspondingauthor{Mouyuan Sun}
\email{msun88@xmu.edu.cn}

\author{Danyang Li}
\affiliation{Department of Astronomy, Xiamen University, Xiamen, Fujian 361005, People's Republic of China; msun88@xmu.edu.cn}

\author[0000-0002-0771-2153]{Mouyuan Sun}
\affiliation{Department of Astronomy, Xiamen University, Xiamen, Fujian 361005, People's Republic of China; msun88@xmu.edu.cn}

\author[0000-0003-4874-0369]{Junfeng Wang}
\affiliation{Department of Astronomy, Xiamen University, Xiamen, Fujian 361005, People's Republic of China; msun88@xmu.edu.cn}

\author[0000-0001-7349-4695]{Jianfeng Wu}
\affiliation{Department of Astronomy, Xiamen University, Xiamen, Fujian 361005, People's Republic of China; msun88@xmu.edu.cn}

\author[0000-0002-2419-6875]{Zhixiang Zhang}
\affiliation{Department of Astronomy, Xiamen University, Xiamen, Fujian 361005, People's Republic of China; msun88@xmu.edu.cn}

\begin{abstract}
We use LCOGT observations (MJD $59434-59600$) with a total exposure time of $\simeq 50$ hours and a median cadence of $0.5$ days to measure the inter-band time delays (with respect to $u$) in the $g$, $r$, and $i$ continua of a highly variable AGN, 6dFGS gJ022550.0-060145. We also calculate the expected time delays of the X-ray reprocessing of a static Shakura \& Sunyaev disk (SSD) according to the sources' luminosity and virial black-hole mass; the two parameters are measured from the optical spectrum of our spectroscopic observation via the Lijiang \SI{2.4}{\meter} telescope. It is found that the ratio of the measured time delays to the predicted ones is $2.6_{-1.3}^{+1.3}$. With optical light curves (MJD $53650-59880$) from our new LCOGT and archival ZTF, Pan-SATRRS, CSS, and ATLAS observations, and infrared (IR) WISE data (MJD $55214-59055$), we also measured time delays between WISE $W1$/$W2$ and the optical emission. $W1$ and $W2$ have time delays (with respect to V), $9.6^{+2.9}_{-1.6}\times 10^2$ days and $1.18^{+0.13}_{-0.10}\times 10^3$ days in the rest-frame, respectively; hence, the dusty torus of 6dFGS gJ022550.0-060145 should be compact. The time delays of $W1$ and $W2$ bands are higher than the dusty torus size-luminosity relationship of~\cite{Lyu2019}. By comparing the IR and optical variability amplitude, we find that the dust covering factors of $W1$ and $W2$ emission regions are 0.7 and 0.6, respectively. Future broad emission-line reverberation mapping of this target and the results of this work enable us to determine the sizes of the AGN main components simultaneously. 
\end{abstract}


\section{Introduction}
\label{sec:intro}
Active galactic nuclei (AGNs) are powered by the accretion gas to their central supermassive black holes \citep[SMBHs; e.g.,][]{Page1974} and emit a continuous spectrum of radiation from extreme ultraviolet (EUV) to infrared (IR) wavelengths \citep[e.g.,][]{Burbidge1967, Shields1978}. Measuring the size and structure of the accretion disk is essential for understanding the SMBH accretion physics and probing the cosmology via gravitationally lensed quasars \citep{Tie2018}. 

In theory, researchers developed several accretion disk models. For instance, the standard geometrically thin but optically thick disk model (SSD; \citealt{Shakura1973}) with a moderate dimensionless accretion rate (i.e., the ratio of the mass accretion rate to the Eddington accretion rate), the optically and geometrically thick disk (SLIM; \citealt{Abramowicz1988}) with a high dimensionless accretion rate, the advection-dominated accretion flow (ADAF) model \citep{Narayan1995} with a low accretion rate, and the luminous hot accretion flow (LHAF) model  \citep{Yuan2007}. It is generally believed that the accretion mode of a luminous AGN is an SSD. 

Due to the small angular size of the AGN central engine, it is challenging to spatially resolve its structure using current facilities. To overcome this limitation, researchers adopt the reverberation-mapping (RM) technique to resolve the central engine by measuring time delays between light curves at different wavelengths (e.g., \citealt{Clavel1991, Cackett2021}). For an SSD, the temperature distribution follows the relation $T(R)\propto R^{-4/3}$, where $T$ and $R$ are the effective temperatures and the disk radius, respectively. Hence, one would expect that the light travel time delay across the SSD obeys the lag-wavelength relation of $\tau(\lambda)\propto \lambda^{4/3}$. However, measuring the time delay of the continuum emission is challenging because of small sizes ($\sim 1$ light day) and weak variations ($\sim 10\%$). Hence, high-cadence observations with high signal-to-noise ratios are required. 

Recent observations (e.g., \citealt{Fausnaugh2016, McHardy2018, Guo_Wei2022, Fian2023, Cackett2023}) reveal that the measured AGN continuum time delays are two or three times larger than the SSD prediction. The discrepancy between the predicted and measured AGN continuum time delays challenges the black-hole accretion theory and is commonly referred to as the ``accretion disk oversize'' problem. 

Several explanations are proposed for the ``accretion disk oversize" problem. For instance, the disk luminosity is underestimated by up to an order of magnitude \citep{Gaskell2017}; the X-ray corona height can be large \citep{Kammoun2021}; the SSD can have strong winds and produce a flat disk temperature profile \citep{Sun2019}; the SSD might emit non-blackbody radiation \citep{Hall2018}; the time-dependent SSD and corona are magnetically coupled \citep{Sun2020}; diffuse emission from the inner parts of the broad-line regions contributes significantly to the observed continua \citep[e.g.,][]{Cackett2018, Sun2018, Cackett2022}; the disk has axisymmetric temperature perturbations \citep{Neustadt2022}; the accretion disk may have a steep rim or rippled structures \citep{Starkey2023}. The universal validity of these explanations has yet to be proven. 

Recent observations of high-luminosity AGNs suggest that the measured time delays are roughly consistent with theoretical expectations \citep{Homayouni2019, Yu2020}. Moreover, \cite{Li2021} found an inverse correlation between the ratio of the measured time delay to the SSD prediction and AGN luminosity (also see \citealt{GuoHengxiao2022}). This correlation is a promising avenue for understanding the physical nature of the ``accretion disk oversize" problem. The anti-correlation should be verified with more disk RM observations. 

The AGN unification model suggests that the central region of an AGN is surrounded by a geometrically and optically thick dusty structure known as the ``dusty torus" \citep[for a recent review, see, e.g.,][]{Netzer2015}. The structure of the dusty torus is unknown, but it is thought to be approximately annular. The continuum emission reaching the dust regions is scattered and absorbed by dust particles and then re-emitted in the IR. The ratio of the AGN's IR luminosity to its optical luminosity ($L_{\mathrm{IR}}/L_{\mathrm{optical}}$) probes the fraction of the solid angle covered by dust, which is the dust covering factor. IR RM allows us to resolve the geometric structure of the dusty torus \citep{Barvainis1992}. 

In this work, we use the disk and IR RM to probe the disk and torus sizes of a highly variable AGN, 6dFGS gJ022550.0-060145. The paper is organized as follows. Section~\ref{sec:2} describes the observations and data. Section~\ref{sec:3} presents the analysis of the time series. Our results are discussed in Section~\ref{sec:4}. We summarize our conclusions in Section~\ref{sec:5}.

Throughout this paper, we adopt the cosmology with $\Omega_{m}=0.3$, $\Omega_{\Lambda}=0.7$, and $H_{0}=70\,\mathrm{km\,s^{-1}\,Mpc^{-1}}$. The reported time delays are in rest-frame unless otherwise specified. 

\section{target and observations}\label{sec:2}
Our target is 6dFGS gJ022550.0-060145 which located at (J2000) RA = 02:25:50.04, DEC = -06:01:45.1 and with a redshift of $z = 0.318$ \citep{Monroe2016}. This source is highly variable according to the Catalina Sky Survey (CSS\footnote{\url{https://catalina.lpl.arizona.edu/}}; \citealt{Drake2009}) and Zwicky Transient Facility (ZTF\footnote{\url{https://www.ztf.caltech.edu/}}; \citealt{Masci2019}) observations (Section~\ref{section:2.1}). We use the Las Cumbres Observatory Global Telescope (LCOGT\footnote{\url{https://lco.global/}}) in four optical bands ($ugri$) from August 3, 2021, to January 16, 2022, to intensively monitor this source. The median cadence of our LCOGT observations is 0.5 days. To probe the intrinsic AGN variability, we require that the signal-to-noise ratios for our observations are higher than $50$ for $g$ ($\sim 4700\,\text{\AA}$), $r$ ($\sim 6215\,\text{\AA}$), and $i$ ($\sim 7545\,\text{\AA}$) bands; for $u$ band ($\sim 3540\,\text{\AA}$), the signal-to-noise is larger than $40$ since its variability amplitude is expected to be higher than other bands. Hence, for the airmass $\leq 1.4$, the required exposure time duration for $u$, $g$, $r$, and $i$ bands are $650$ s, $90$ s, $60$ s, and $60$ s, respectively. The total LCOGT exposure time is $\sim 50$ hours. 

\subsection{Multi-band Light curves}\label{section:2.1}

\begin{figure*}
\centering
\plotone{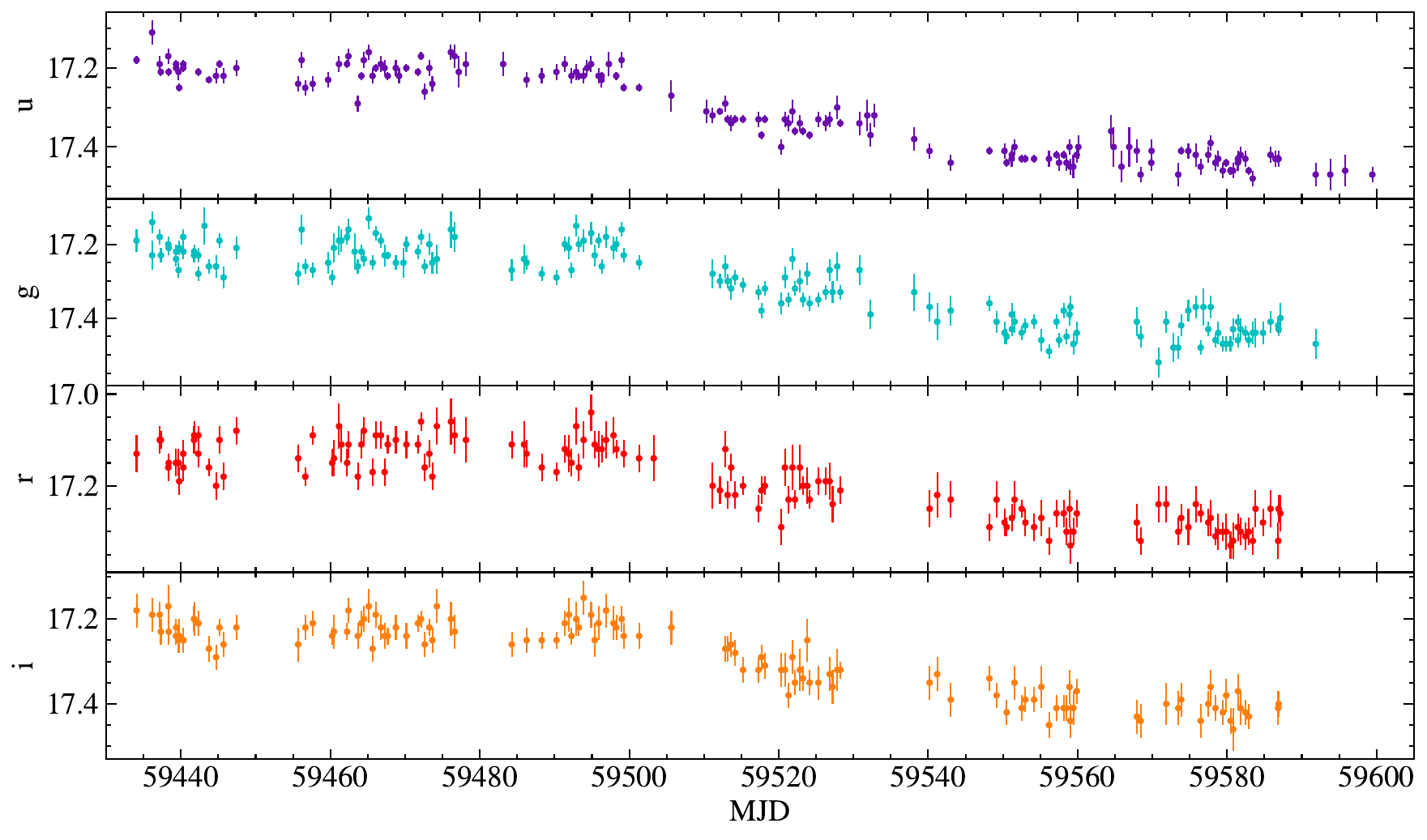}
\caption{The light curves of 6dFGS gJ022550.0-060145, observed by LCOGT $u$, $g$, $r$, and $i$ bands.}
\label{fig:LCO_lc}
\end{figure*}

We use the Python3-based Automated Photometry Of Transients \citep[AutoPhOT\footnote{\url{https://github.com/Astro-Sean/autophot}};][]{Brennan2022} package to perform PSF photometry on the LCOGT images of 6dFGS gJ022550.0-060145. AutoPhot is a convenient automated pipeline to estimate the image point spread function (PSF) and perform the PSF fitting and the zero-point calibration. We calibrate the $g$, $r$, $i$, and $u$ magnitudes against the custom catalog from the $16$th Data Release of the Sloan Digital Sky Surveys\footnote{\url{https://skyserver.sdss.org/dr16/en/home.aspx}} \citep{Ahumada2020}. Full details of all-optical photometric measurements determined with AutoPhOT are given in \hyperref[sec:appendix]{Appendix}. The light curves are shown in Figure~\ref{fig:LCO_lc} and given in Table~\ref{table: lcolc}. 

\begin{deluxetable}{cccc}
\tablecaption{LCOGT Light Curves \label{table: lcolc}}
\tablehead{\colhead{Filter}& \colhead{MJD (J2000)} & \colhead{Mag} & \colhead{Magerr} } 
\startdata
$u$  & 59434.0581 & 17.18 & 0.01 \\
{} & 59436.1462 & 17.11 & 0.03 \\
{} & 59437.1560 & 17.19 & 0.02 \\
{} &   \dots     &   \dots  &  \dots  \\
$g$  & 59434.0648 & 17.19 & 0.03 \\
{} & 59436.1529 & 17.14 & 0.03 \\
{} & 59436.1532 & 17.23 & 0.04 \\
{} &   \dots     &   \dots  &  \dots  \\
$r$  & 59434.0657 & 17.13 & 0.04 \\
{} & 59437.1638 & 17.10 & 0.03 \\
{} & 59437.3134 & 17.10 & 0.02 \\
{} &   \dots     &   \dots  &  \dots  \\
$i$  & 59434.0665 & 17.18 & 0.04 \\
{} & 59436.1546 & 17.19 & 0.04 \\
{} & 59437.1648 & 17.19 & 0.03 \\
\enddata
\tablecomments{This table is available in its entirety in the online version of this manuscript.}
\end{deluxetable}

\begin{figure*}
\plotone{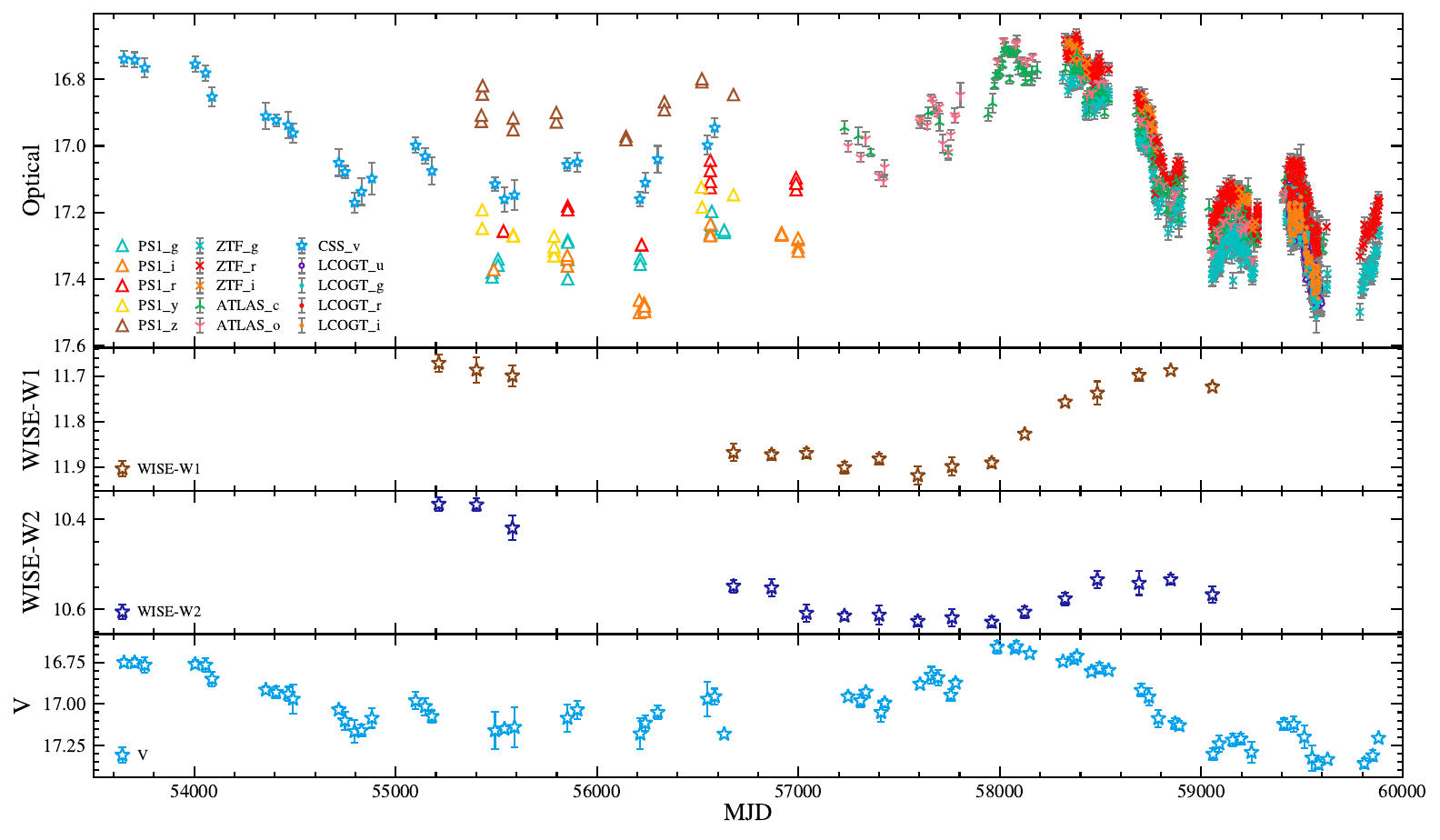}
\caption{6dFGS GJ022550.0-060145's multi-band light curves. Light curves in the top panel are from PS1, ATLAS, CSS, ZTF, and LCOGT. The second and third panels show the WISE $W1$ and $W2$ light curves. The bottom panel presents the rebinned synthetic $V$-band light curve. }
\label{fig:light curve}
\end{figure*}

We check the archival data of our target. ZTF observed our target in $g$ ($\sim 4784\,\text{\AA}$), $r$ ($\sim 6417\,\text{\AA}$), and $i$ ($\sim 7867\,\text{\AA}$) bands, Pan-STARRS (PS1\footnote{\url{https://outerspace.stsci.edu/display/PANSTARRS/}}; \citealt{Flewelling2020}) in $g$ ($\sim 4866\,\text{\AA}$), $r$ ($\sim 6215\,\text{\AA}$), $i$ ($\sim 7545\,\text{\AA}$), and $zs$ ($\sim 8679\,\text{\AA}$) bands, CSS in $V$($\sim 5550\,\text{\AA}$) band, and Asteroid Terrestrial-impact Last Alert System (ATLAS\footnote{\url{https://fallingstar-data.com/forcedphot/}}; \citealt{Tonry2018, Smith2020}) in $o$ ($\sim 6780\,\text{\AA}$) and $c$ ($\sim 5330\,\text{\AA}$) bands. The optical light curves were compiled and shown in the upper panel of Figure~\ref{fig:light curve}. It is obvious that 6dFGS GJ022550.0-060145 has several optical flares: for instance, the source is dimmed by about 0.4 magnitudes from MJD $54000$ to $56000$, brightened by about 0.5 magnitudes from MJD $56000$ to $58000$, and dimmed again by about 0.6 magnitudes from MJD $58000$ to $59100$. Comparing the $g$-$r$ colors in the low (e.g., MJD$=56000$, $59000$) and high (around MJD$=58400$) flux states, it is evident that the AGN become optically bluer as the flux level increases. This bluer-when-brighter behaviour is observed in quasars \citep[e.g.,][]{Schmidt2012, Sun2014}.

We have g and r band light curves from LCOGT and ZTF. To merge them into new g and r band light curves and improve the cadences, we performed inter-calibrations to ensure light curves from different telescopes were on the same flux scale. Our inter-calibration process was as follows. First, we calculated the difference between the mean LCOGT and ZTF $g$ ($r$) magnitudes, i.e., $\overline{{mag}_{\mathrm{lco}}}- \overline{{mag}_{\mathrm{ztf}}}$ 
. Note that we only considered ZTF observations with $\mathrm{MJD>59400}$ and $\mathrm{MJD<59650}$ for the inter-calibration. Second, the difference was added to the ZTF $g$ ($r$) light curves, and the results were "appended" to the LCOGT $g$ ($r$) observations. 

Our target was also observed by the Wide-field IR Survey Explorer (WISE\footnote{\url{https://www.nasa.gov/mission_pages/WISE/main/index.html}}; \citealt{Wright2010}) and the Near-Earth Object WISE Reactivation mission (NEOWISE\footnote{\url{https://www.nasa.gov/mission_pages/neowise/main/index.html}}; \citealt{Mainzer2011}) in $W1$ and $W2$ bands.  WISE scanned the sky every six months except in the hibernation phase (from February 1st, 2011 to October 3rd, 2013). We reject low-quality WISE observations with the flags ``qi\_fact'' $<1$, ``SSA'' <5, and ``moon\_mask'' $=1$. The remaining data points are rebinned by every $180$ day; in the rebin, we use the median value whose uncertainty is estimated via the average absolute deviation. The rebinned light curves are shown in Figure~\ref{fig:light curve}. Like optical observations, WISE $W1$ and $W2$ light curves show evident mid-IR variations. 

To compare the optical long-term light curves with the extensive yet sporadic WISE data, we construct a long synthetic $V$ band light curve from the LCOGT, ZTF, CSS, PS1, and ATLAS data. First, we convert the ATLAS $o$ and $c$ magnitudes into $g$ and $r$ magnitudes following Equation~\ref{eq3} \citep{Tonry2018}:
\begin{equation} \label{eq3}
  \begin{cases}
     g=1.67*c-0.67*o \\
     r=0.35*c+0.65*o.
   \end{cases}
\end{equation}
Second, we convert all $g$ and $r$ magnitudes into the synthetic $V$-band magnitudes by using Equation~\ref{eq1}~\citep{Jester2005}:
\begin{equation} \label{eq1}
V=g-0.52*(g-r)-0.03.
\end{equation}
Third, we rebin the synthetic $V$-band magnitudes every 50 days (the bottom panel of Figure~\ref{fig:light curve}). The rebinned $V$-band light curve as used to determine the correlation coefficients and time delays between $V$ and WISE bands. Both the WISE and $V$-band light curves are irregularly sampled and interpolations are required in the cross-correlation analysis (see Section~\ref{sec:3.1}). We use the rebinned light curves rather than the raw data to ensure that the interpolation is robust and not sensitive to possible outliers in the raw data.

\subsection{Optical spectrum}\label{sec:2.2}
We aim to use the SSD to predict the time delays of 6dFGS gJ022550.0-060145 and compare the predicted with the measured ones. To do so, we need to measure the black hole mass ($M_{\mathrm{BH}}$) of our target. We use the Yunnan Faint Object Spectrograph and Camera (YFOSC) mounted on the Lijiang \SI{2.4}{\meter} telescope at the Yunnan Observatories of the Chinese Academy of Sciences to take a high signal-to-noise ratio optical spectrum of our source on Nov 20, 2022. The YFOSC has a 2k$\times$4k back-illuminated CCD detector \citep{Lu2019, Wang2019-LJ}. We adopt the ``grism8'' configuration with a 2.5 arcsec slit. The corresponding wavelength coverage and resolution are $3858-7330\,\text{\AA}$ and $4900$, respectively. The exposure time is 30 minutes, and the corresponding spectral signal-to-noise ratio is 19. We employed the \texttt{IRAF V2.17} software to perform the spectroscopic reduction. The reduction process includes bias correction, flat correction, spectral extraction, and wavelength calibration. By rotating the long slit, we exposed the object star with a nearby comparison star simultaneously. The comparison star is used as the standard to do the flux calibration and telluric correction (for more details, please see \citealt{lu2021}). 

\begin{figure*}
\centering
\plotone{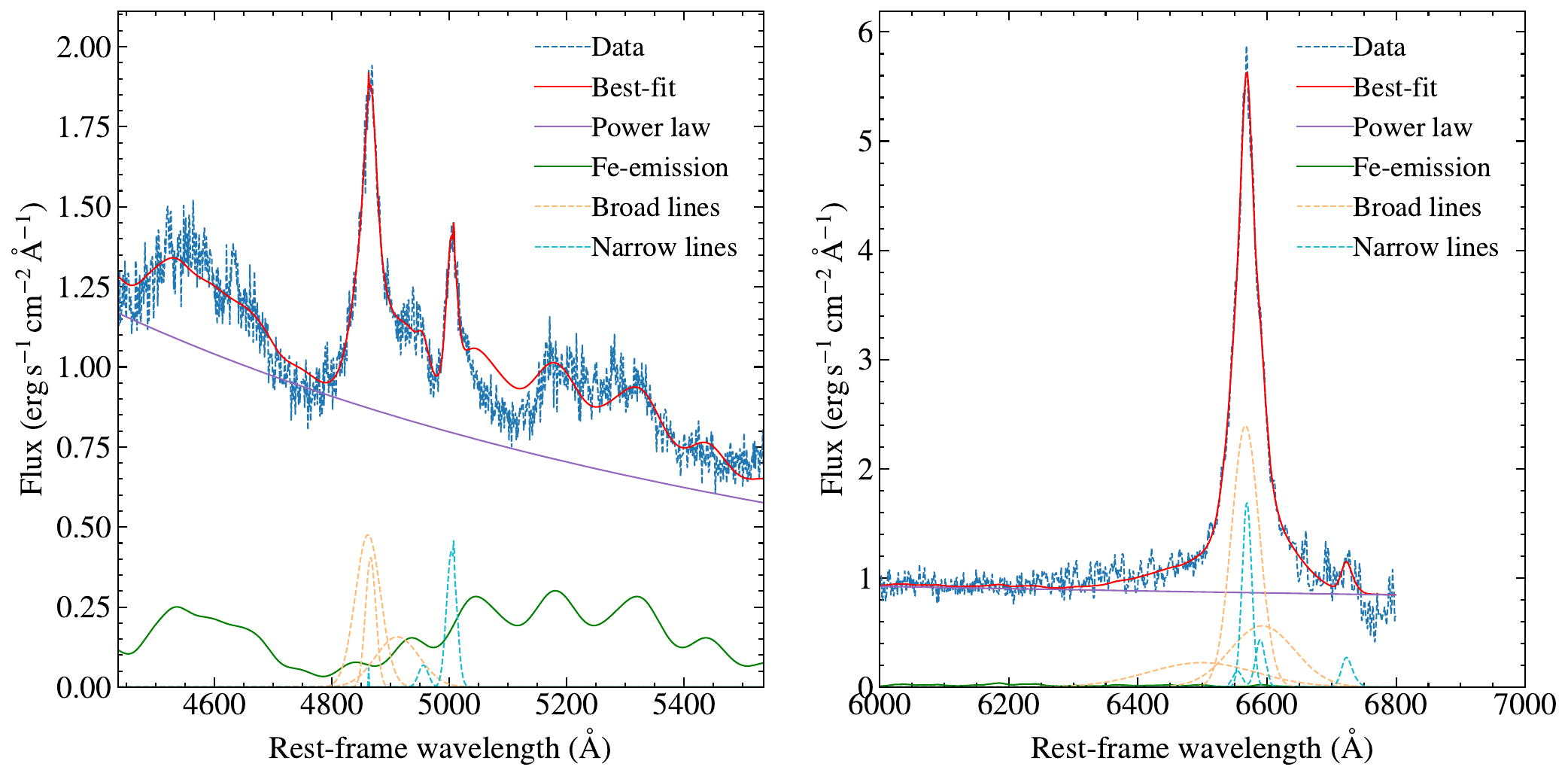}
\caption{The spectroscopic decomposition results for 6dFGS gJ022550.0-060145. The left and right panels are for H$\alpha$ and H$\beta$, respectively. The dark blue dashed curves and red solid curves represent the data and best-fitting models, respectively. The purple and green curves are for the power-law continua and the best-fitting iron templates. The light blue dashed and yellow curves correspond to the narrow and broad emission lines, respectively.}
\label{fig:spetrum}
\end{figure*}

We perform the quasar spectral decomposition analyses to the Lijiang 2.4m spectrum following our previous work \citep[e.g.,][]{Sun2015}. The adopted continuum model consists of a power law and the iron template of \cite{Vestergaard2001}. Three Gaussians are used to fit the continuum-subtracted broad emission lines. As for narrow emission lines, we use a single Gaussian to fit each one. The best-fitting model is obtained by minimizing the $\chi^2$ statistic. The Full Width at Half Maximum (FWHM) of each best-fitting broad emission-line profile (a summation of the three Gaussian components) is measured. The continuum luminosity at rest-frame $5100\,\text{\AA}$ ($5100\,\text{\AA}L_{\mathrm{5100\,\text{\AA}}}$) and the H$\alpha$ luminosity ($L_{\mathrm{H\alpha}}$) is also calculated from the best-fitting power-law model and the H$\alpha$ broad emission-line profile, respectively. Figure~\ref{fig:spetrum} shows our optical spectrum fitting results. Some best-fitting parameters are shown in Table~\ref{table: spectrum}. Then, We use the H$\alpha$ and H$\beta$ virial mass estimators to estimate $M_{\mathrm{BH}}$ (i.e.,  Equations 5 and 10 in \citealt{Shen2011}):
\begin{equation} \label{eq4} 
  \begin{cases}
  
     \mathrm{log}(\frac{M_{\mathrm{BH}}}{M_{\odot}}) = a_{1}+b_{1}\,\mathrm{log}(\frac{L_{\mathrm{H\alpha}}}{\mathrm{10^{42}\,erg\,s^{-1}}})+c_{1}\,\mathrm{log}(\frac{\mathrm{FWHM}_{\mathrm{H\alpha}}}{\mathrm{km\,s^{-1}}}) \\
     
     \mathrm{log}(\frac{M_{\mathrm{BH}}}{M_{\odot}})=a_{2}+b_{2}\,\mathrm{log}(\frac{5100\,\text{\AA}\,L_{\mathrm{5100\,\text{\AA}}}}{\mathrm{10^{44}\,erg\,s^{-1}}})+c_{2}\,\mathrm{log}(\frac{\mathrm{FWHM}_{\mathrm{H\beta}}}{\mathrm{km\,s^{-1}}}),
     
   \end{cases}
\end{equation}
where $M_{\odot}$ is the solar mass, the coefficients $a_1$, $a_2$, $b_1$, $b_2$, $c_1$, and $c_2$ are 0.379, 0.91, 0.43, 0.5, 2.1, and 2, respectively. The $\log (M_{\mathrm{BH}}/M_{\odot})$ for H$\alpha$ and H$\beta$ are $8.11_{-0.03}^{+0.03}$ and $7.89_{-0.04}^{+0.04}$, respectively. We take the average of the two as our fiducial $\log (M_{\mathrm{BH}}/M_{\odot})=8.02_{-0.02}^{+0.02}$. 

We cannot use the RM to obtain the size of the BLR because of the lack of broad emission-line light curves. Early reverberation mapping studies suggest that the H$\beta$ BLR size ($R_{\mathrm{H}\beta}$) correlates tightly with $5100\,\text{\AA}L_{\mathrm{5100\,\text{\AA}}}$ \citep{Bentz2009}. Recently, it has been found that the BLR radius depends upon both the AGN luminosity and the Eddington ratio \cite[e.g.,][]{Du2016, Du2018, Fonseca2020}, i.e., the BLR radius-luminosity relation of \cite{Bentz2009} can significantly over-estimate the true BLR size for high-Eddington ratio sources. The Eddington ratio of our target is low because $L_{\mathrm{AGN}}/L_{\mathrm{Edd}}= 0.23\pm 0.01$, where $L_{\mathrm{AGN}}$ is the bolometric luminosity (see Eq.~\ref{eq6}) and $L_{\mathrm{Edd}}$ is the Eddington luminosity. Therefore, we use the BLR radius-luminosity relation (i.e., Equation 2 in \citealt{Bentz2009}) to roughly estimate $R_{\mathrm{H}\beta}$, which is $83.5_{-0.6}^{+0.5}$ light days.

\begin{deluxetable}{ccc}
\tablecaption{Spectral fitting parameters}
\label{table: spectrum}
\tablehead{\colhead{Broad emission line} & \colhead{Parameter} & \colhead{Value} } 
\startdata
H$\beta$ & $\mathrm{log}(5100\,\text{\AA}\,L_{\mathrm{5100\,\text{\AA}}}/[\,\mathrm{erg\,s^{-1}}])$& $44.612_{-0.004}^{+0.003}$\\
{} & $\mathrm{FWHM}_{\mathrm{H\beta}}/[\,\mathrm{km\,s^{-1}}]$ & $2.18_{-0.09}^{+0.12}\times10^{3}$ \\
{} & $\mathrm{log}(M_{\mathrm{BH}}/\,M_{\odot})$	& $7.89_{-0.04}^{+0.04}$ \\
\hline
H$\alpha$ & $\mathrm{log}(L_{\mathrm{H\alpha}}/[\,\mathrm{erg\,s^{-1}}])$& $43.23_{-0.03}^{+0.04}$\\
{} & $\mathrm{FWHM}_{\mathrm{H\alpha}}/[\,\mathrm{km\,s^{-1}}]$ & $2.69_{-0.07}^{+0.09}\times 10^{3}$ \\
{} & $\mathrm{log}(M_{\mathrm{BH}}/\,M_{\odot})$	& $8.11_{-0.03}^{+0.03}$ 
\enddata
\end{deluxetable}

\section{Time series analysis}\label{sec:3}

\subsection{Time delay}
\label{sec:3.1}
\begin{figure*}
\epsscale{2}
\plottwo{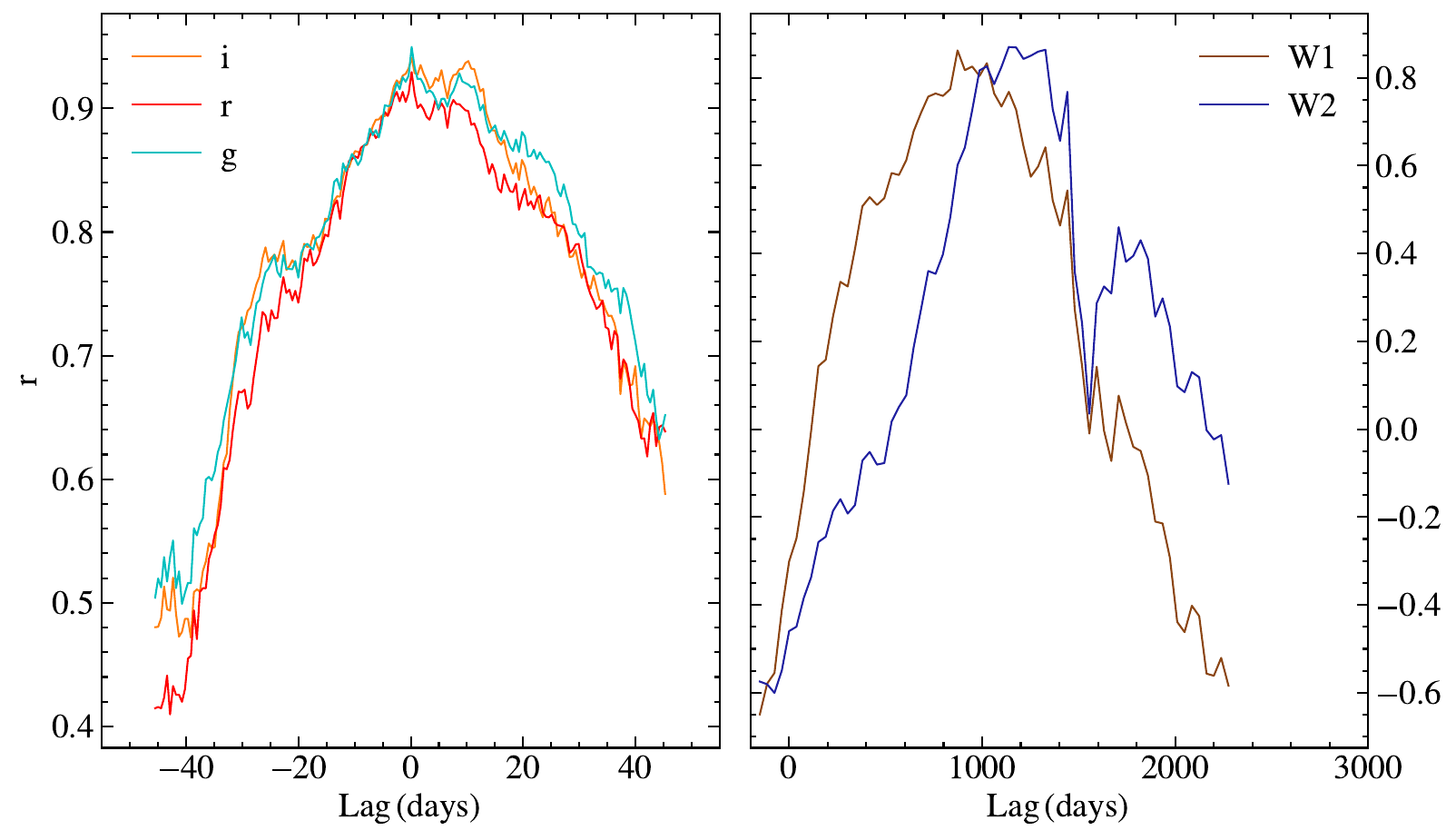}{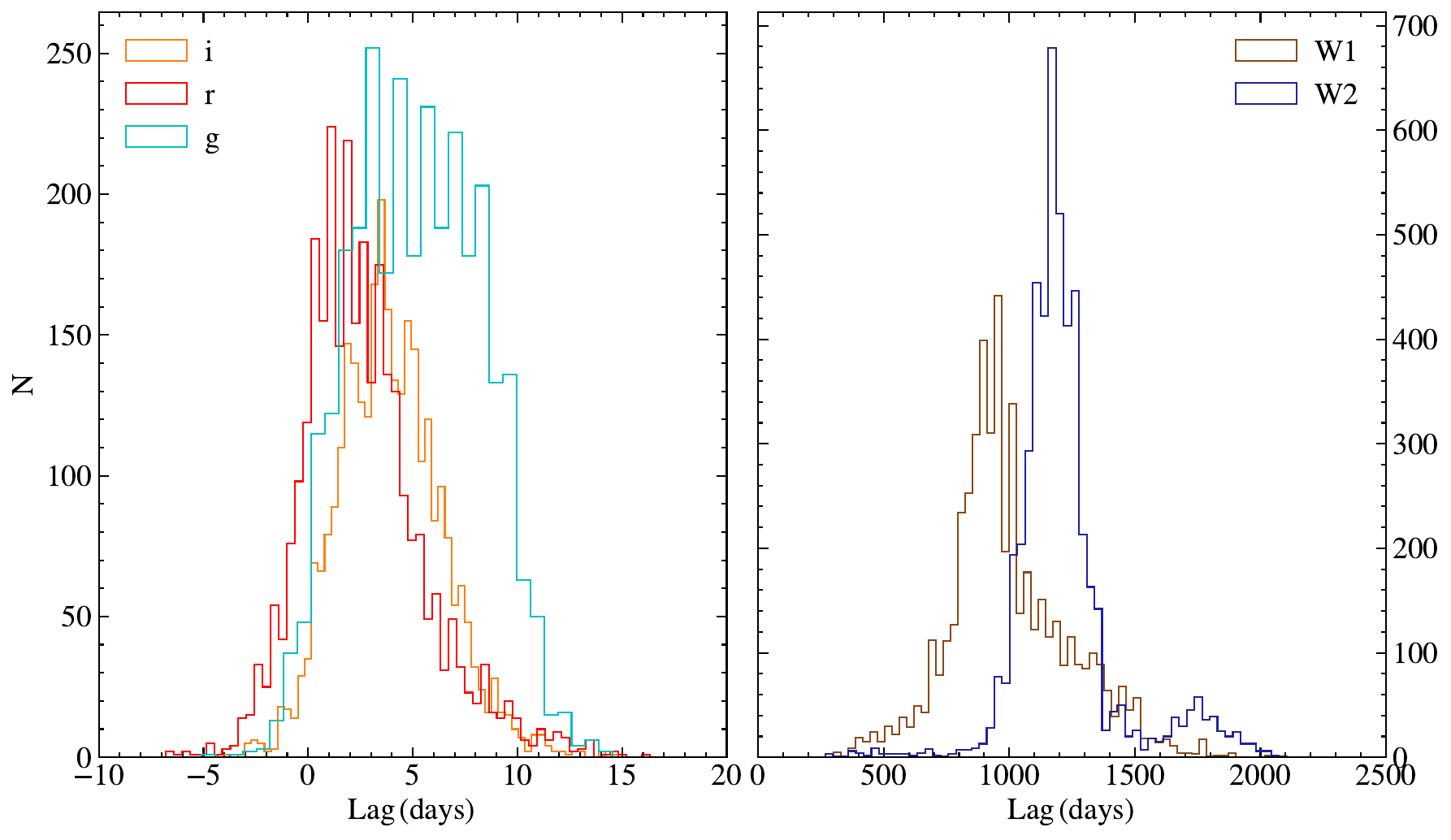}
\caption{The \emph{PyCCF} time delay measurements. The upper-left and upper-right panels are the optical and infrared CCFs. The lower-left panel shows the rest-frame time delays of $g$, $r$, and $i$ (with respect to $u$). The upper-right panel demonstrate the time delays of WISE $W1$ and $W2$ (with respect to $V$).} 
\label{fig:ccf}
\end{figure*}

To estimate time delays between each band relative to the $u$-band for 6dFGS GJ022550.0-060145, we use \emph{PyCCF}~\citep{Sun2018_ccf} which is a Python tool to obtain the interpolated cross-correlation coefficient ($r$) as a function of the time lag. The lag ranges of \emph{PyCCF} are from $-60$ days to $60$ days (which are significantly longer than our expected time delays, i.e., several days), with a uniform step of $0.7$ days (i.e., roughly 1.5 times the median cadence). The measured lags are estimated from the centroids of the interpolated cross-correlation functions (ICCFs), i.e., the $r$-weighted mean lags whose $r>0.9r_{\mathrm{max}}$. \emph{PyCCF} employs the random subset selection and flux redistribution to assess the underlying distribution of the time delay. The ICCFs and time-delay distributions are shown in the upper-left and lower-left panels of Figure~\ref{fig:ccf}. The $50$-th, $84$-th, and $16$-th percentiles of the distributions are the measured time delay and the $1\sigma$ upper and lower limits. The time delay results of the $g$, $r$, and $i$ bands (with respect to $u$) are $5.2_{-3.2}^{+3.3}$ days, $2.3_{-2.2}^{+3.0}$ days, and $3.8_{-2.3}^{+2.5}$ days, respectively. 

The dusty torus produces IR emissions. Therefore, measuring the time delays of the $W1$ and $W2$ (with respect to the optical emission) can probe the sizes of the dusty torus. We again use \emph{PyCCF} to measure the time delay of the WISE $W1$ (or $W2$) light curve with respect to the $V$-band light curve. The lag ranges of \emph{PyCCF} are from $-200$ days to $3000$ days, with a uniform step of $50$ days. We use the asymmetric lag ranges because, physically speaking, the dust emission should respond to optical variations with significant time delays. The CCFs and time delay distributions are shown in the upper-right and lower-right panels of Figure~\ref{fig:ccf}. The results are $9.6^{+2.9}_{-1.6}\times 10^2$ days (for $W1$) and $1.18^{+0.13}_{-0.10}\times 10^3$ days (for $W2$). That is, the distances of $W1$ and $W2$ to the V band are $9.6^{+2.9}_{-1.6}\times 10^2$ light days and $1.18^{+0.13}_{-0.10}\times 10^3$ light days, respectively. Hence, the BLR radius inferred from the BLR radius-luminosity relation is smaller than the dust reverberation radius by a factor of $11\sim 14$. This factor is larger than those in \cite{Koshida2014}. In addition, the emission regions of $W1$ and $W2$ are statistically identical to each other. 

\subsection{The damping timescale}

Quasar UV/optical light curves are often fitted by the damped random walk (DRW) model \citep[e.g.,][]{Kelly2009, Sun2018, Burke2021}. The covariance matrix of the DRW model is $S=\sigma^2 \exp{(-|\Delta t|/t_{\mathrm{damping}}})$, where $\Delta t$, $\sigma$, and $t_{\mathrm{damping}}$ are the time interval between two observations, the short-term variability amplitude, and the damping timescale. It is often speculated that the damping timescale is closely related to quasar properties \citep[e.g.,][]{Kelly2009, Sun2018}. For instance, \cite{Burke2021} obtain the relationship between the DRW damping timescale and SMBH mass: 
\begin{equation} \label{eq55}
t_{\mathrm{damping}} = 107_{-12}^{+11}\,\mathrm{days}\,(\frac{M_{\mathrm{BH}}}{10^{8}\,M_{\odot}})^{0.38_{-0.04}^{+0.05}}.
\end{equation}
Meanwhile, it is stressed that the damping timescale can easily be biased unless the intrinsic $t_{\mathrm{damping}}$ is less than $10\%$ of the light-curve duration \citep[e.g.,][]{Kozlowski2017, Hu2023}. The synthetic $V$-band light curve of our target has a long duration of $\sim 6230$ days (observed frame), enabling us to robustly measure the damping timescale since the expected damping timescale in the observed frame is only $109(1+z)=143.7$ days (Equation~\ref{eq55}; \citealt{Burke2021}). We fit the synthetic $V$-band light curve with the DRW model via \emph{JAVELIN}\footnote{\url{https://github.com/nye17/javelin}} \citep{Zu2011,Zu2013}. We search the best-fitting DRW parameters by the Markov Chain Monte–Carlo (MCMC) sampling and obtain the amplitude $\sigma=0.19_{-0.04}^{+0.06}$ mag and the rest-frame damping timescale $t_{\mathrm{damping, obs}}= 7.5^{+7.1}_{-2.9}\times 10^2$ days. Our measured damping timescale is about $7$ times larger than the expected value. Hence, it might be that the damping timescale predicted by Equation~\ref{eq55} \citep{Burke2021} are under-estimated (S. Zhou et al. 2024, in preparation).  

\section{DISCUSSION}\label{sec:4}

In this section, we discuss the accretion disk and dusty torus sizes of 6dFGS GJ022550.0-060145 according to the measured time delays. 

\subsection{Accretion disk size}
We estimate the bolometric luminosity ($L_{\mathrm{AGN}}$) from $5100\,\text{\AA}\, L_{5100\,\text{\AA}}$ with the quasar bolometric correction of \cite{Runnoe2012}:

\begin{equation} \label{eq6}
\mathrm{log}(\frac{L_{\mathrm{AGN}}}{\mathrm {erg\,s^{-1}}})=4.89+\mathrm{0.91\,log}(\frac{5100\,\text{\AA}\,L_{\mathrm{5100\,\text{\AA}}}}{\mathrm{erg\,s^{-1}}}),
\end{equation}
and the result is $\mathrm{log}(L_{\mathrm{AGN}}/[\mathrm {erg\,s^{-1}}])=45.49$. According to the SSD, the expected time delay ($\tau_{\mathrm{SSD}}$; with respect to the $u$ band) of 6dFGS gJ022550.0-060145 as a function of the rest-frame wavelength is (e.g., \citealt{Fausnaugh2016, Li2021})
\begin{equation} \label{eq5}
     \frac{\tau_{\mathrm{SSD}}}{\mathrm{1.0\,days}}=\tau_{\mathrm{0, SSD}}((\frac{\lambda}{\lambda_{u}})^{4/3}-1),
\end{equation}
where $\tau_{\mathrm{0, SSD}}=0.65(\frac{M_{\mathrm{BH}}}{10^{8}\,M_{\odot}})^{2/3}(\frac{\dot{m}}{0.1})^{1/3}(\frac{\lambda_{u}}{2500\,\mathrm{\text{\AA}}})^{4/3}=0.958$ days, 
$\dot{m}=0.23^{+0.01}_{-0.01}$ is the ratio of $L_{\mathrm{AGN}}$ to the Eddington luminosity, and $\lambda_{u}$ is the rest-frame $u$-band wavelength. Note that this formula is valid if the radiative efﬁciency is $\eta = 0.1$. 

\begin{figure}
\centering
\epsscale{1}
\plotone{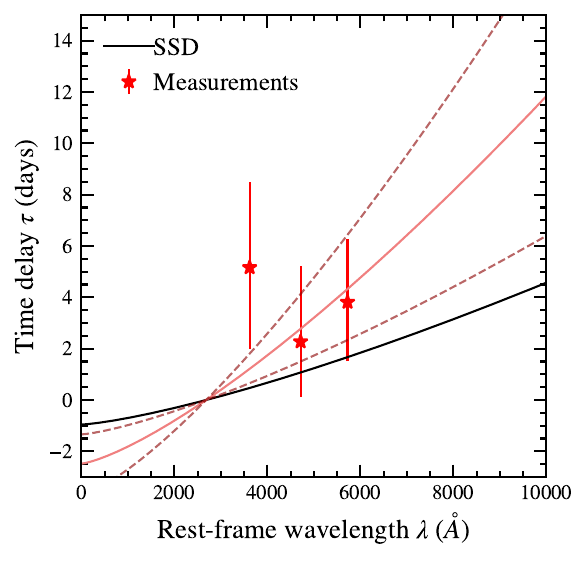}
\caption{\label{fig:lag_spectrum}Time delays (with respect to the $u$ band) as a function of wavelengths in rest-frame. The black curve shows the SSD predictions for 6dFGS gJ022550.0-060145. The red solid curve is the best-fitting relationship; the red dashed curves correspond to the $1\sigma$ confidence intervals.} 
\end{figure}

We compare the measured time delays (red stars) against the SSD predictions (the black curve) in Figure~\ref{fig:lag_spectrum}. We use \emph{emcee}\footnote{\url{https://github.com/dfm/emcee}} \citep{Foreman-Mackey2013} to fit Equation~\ref{eq_tue} to the rest-frame time delay measurements of the $g$ ($\sim 3566\,\text{\AA}$ in rest-frame), $r$ ($\sim 4715\,\text{\AA}$ in rest-frame), and $i$ ($\sim 5725\,\text{\AA}$ in rest-frame) bands relative to the $u$ band ($\sim 2686\,\text{\AA}$ in rest-frame), which are $5.2_{-3.2}^{+3.3}$ days, $2.3_{-2.2}^{+3.0}$ days, and $3.8_{-2.3}^{+2.5}$ days (Section~\ref{sec:3.1}), respectively. 

\begin{equation} \label{eq_tue}
     \tau=\tau_{0}((\frac{\lambda}{\lambda_{u}})^{4/3}-1),
\end{equation}
where $\tau_0$ is a free parameter, $\lambda$ is the rest-frame wavelength, and $\lambda_{u}$ is the rest-frame wavelength of the $u$ band. The fitting method is via the maximum likelihood method, and the likelihood is $\ln \mathcal{L}=-0.5\sum [(\tau_{\mathrm{obs}}-\tau_{\mathrm{th}})^2/\sigma_{\tau}^2 + \ln (2\pi \sigma_{\tau}^2)]$, where $\tau_{\mathrm{obs}}$ is the measured rest-frame time lag, $\sigma_{\tau}$ is the $1\sigma$ uncertainty of $\tau_{\mathrm{obs}}$, and $\tau_{\mathrm{th}}$ is given by Equation~\ref{eq_tue}. The best-fit value of $\tau_{0}$ is $2.5_{-1.1}^{+1.2}$, and the best-fitting relation is shown as the red curve in Figure~\ref{fig:lag_spectrum}. Comparing $\tau_0$ with $\tau_{\mathrm{0, SSD}}$, we find that the ratio of the former to the latter is $2.6^{+1.3}_{-1.3}$(the uncertainties are estimated via MCMC). This ratio seems larger than unity, albeit with substantial uncertainties. Many previous works also suggest the ``accretion disk oversize'' problem (e.g., \citealt{Fausnaugh2016, McHardy2018, Guo_Wei2022, Fian2023, Sharp2023}), i.e., the observed time lags are about three times larger than the SSD prediction. Moreover, \cite{Li2021} proposes that this ratio of the observed to SSD time lag inversely correlates with the AGN luminosity \citep[see also][]{GuoHengxiao2022}. Given its larger uncertainties, 6dFGS gJ022550.0-060145 fits the anti-correlation found by \cite{Li2021} and \cite{GuoHengxiao2022}. Future better time-lag measurements of this source can test the anti-correlation and provide additional clues to the ``accretion disk oversize'' problem. 

Another notable feature is that the time lag of $g$ (with respect to $u$) seems to be larger than the time lags of $r$ and $i$ (with respect to $u$). Suppose this feature is real rather than due to statistical uncertainties. In that case, the feature might be related to the lag excess in the Balmer jump (whose rest-frame wavelength is $\sim 3600\ \text{\AA}$) as seen in some local AGNs \citep[e.g.,][and references therein]{Cackett2018, Cackett2023}. 

Inter-band time lags are often interpreted as the X-ray light travel time differences across different disk emission regions. In this idea, the variable X-ray emission can illuminate the accretion-disk surface and is reprocessed as variable UV/optical emission \citep[e.g.,][]{Krolik1991, Cackett2007}. If so, the X-ray luminosity should be comparable to the disk power for our highly variable target \citep[a similar argument is made by][for a changing-look quasar]{Dexter2019}. We estimate the rest-frame monochromatic luminosity of the $u$-band using the average magnitude ($\sim17.315$) of our LCOGT observations, which is $L_{u} = 10^{30.03}\ \mathrm{erg\,s^{-1}\, Hz^{-1}}$; at the redshift of $0.318$, the rest-frame effective wavelength for the $u$-band is $2686\ \text{\AA}$. Hence, we expect the rest-frame monochromatic luminosity at $2500\ \text{\AA}$, $L_{2500\,\text{\AA}}\simeq L_{u}$. Then, we estimate the rest-frame monochromatic luminosity at $2$ keV, $L_{\mathrm{2keV}}$, via the well-known empirical relation between $L_{\mathrm{2keV}}$ and $L_{2500\,\text{\AA}}$ \citep[e.g.,][]{Steffen2006}:
\begin{equation} \label{eq_2kev}
    \mathrm{log}(\frac{L_{\mathrm{2keV}}}{\mathrm{erg\,s^{-1}\,Hz^{-1}}})=0.642\,\mathrm{log}(\frac{L_{2500\,\text{\AA}}}{\mathrm{erg\,s^{-1}\,Hz^{-1}}})+6.873 \\,
\end{equation}
i.e., $L_{\mathrm{2keV}}= 10^{26.15}\ \mathrm{erg\,s^{-1}\, Hz^{-1}}$. For the power-law spectrum with a photon index of $-2$, the total X-ray luminosity $L_{\mathrm{X}}\simeq \mathrm{2keV} L_{\mathrm{2keV}}=7\times 10^{43}\ \mathrm{erg\,s^{-1}}$. The ratio of $L_{\mathrm{X}}$ to $L_{\mathrm{AGN}}$ (the bolometric luminosity) is only $2.3\%$. Hence, the observed X-ray emission is not powerful enough to drive the observed optical variations unless the corona's X-ray emission is extremely anisotropic. The same conclusion is obtained by \cite{Dexter2019} and \cite{Marcin2023}, who obtain inter-band time lags for an X-ray weak quasar. The recent three-dimensional magnetohydrodynamic simulations of AGN accretion disk also suggest that X-ray is unlikely to be the main driver of UV/optical variations \citep{Secunda2023}. The UV reprocessing \citep{Gardner2017} or magnetic coupling between the corona and disk \citep{Sun2020} may play important roles in driving UV/optical coordinated variations with inter-band lags. 

\subsection{The Dusty Torus of 6dFGS gJ022550.0-060145}
\begin{figure*}
\centering
\plotone{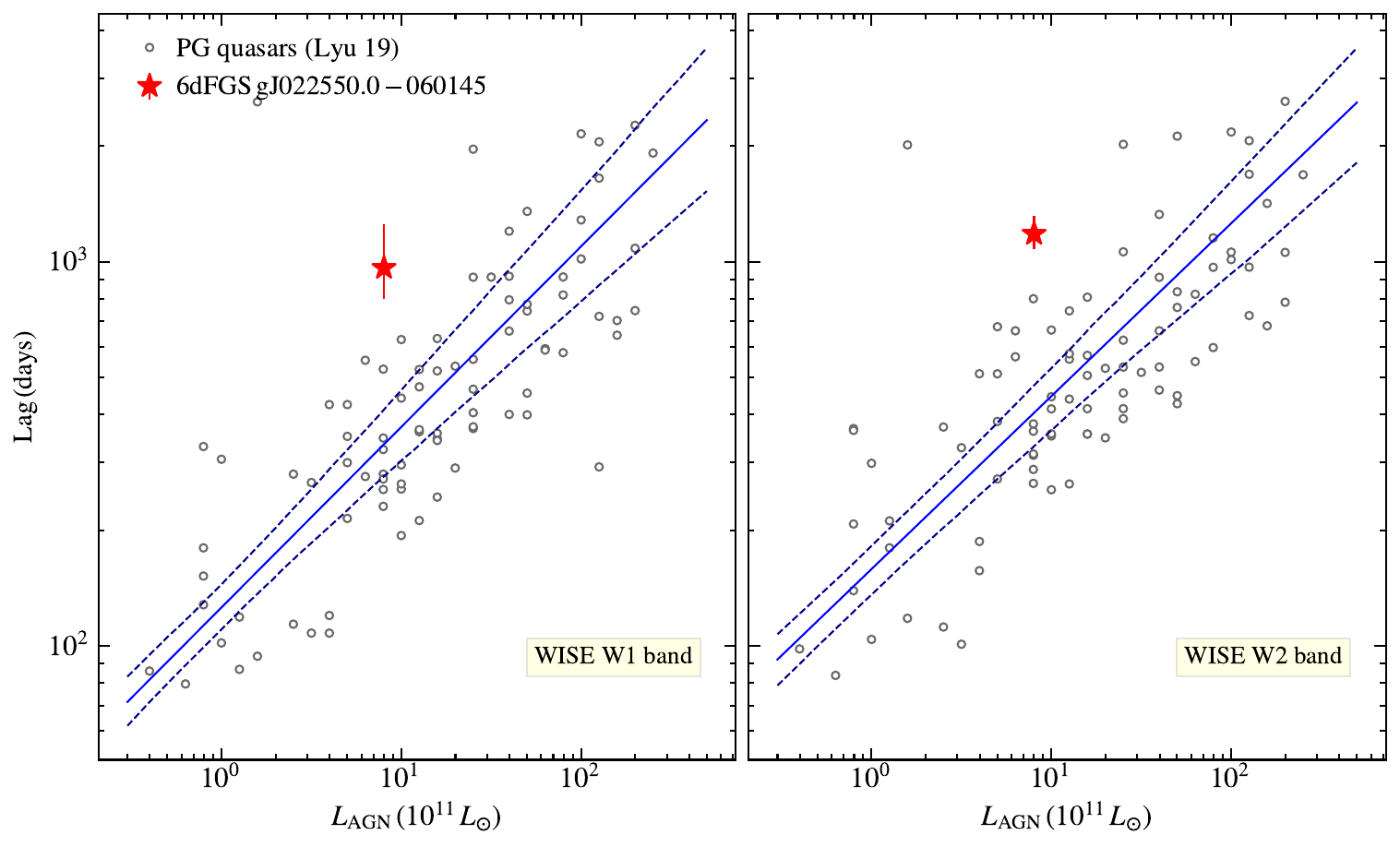}
\caption{\label{fig:L-lag}The rest-frame time delay between $W1$ ($W2$) and the optical emission. The left and right panels are for $W1$ and $W2$, respectively. In each panel, the grey dots are the PG quasars of \cite{Lyu2019}, and the blue lines are their best-fitting relations and $1 \sigma$ confidence ranges. The red star point represents 6dFGS GJ022550.0-060145.}
\end{figure*}

We present our target on the infrared time delay-$L_{\mathrm{AGN}}$ plane (see Figure~\ref{fig:L-lag}) of \cite{Lyu2019}. Comparing with the infrared time delay-$L_{\mathrm{AGN}}$ of \cite{Lyu2019}, 6dFGS gJ022550.0-060145 have larger time delays in $W1$ (by a factor of $2.5_{-0.3}^{+0.2}$) and $W2$ (by a factor of $3.5_{-0.4}^{+0.3}$) bands. Similar to \cite{Lyu2019}, we also find the time delay ratio between $W1$ and $W2$ is $0.8 _{-0.5}^{+0.7}$, indicating that the torus structure is compact. 

We can use the $V$-band light curve and the time delay-corrected WISE light curves to infer the torus covering factor, which is defined as  
\begin{equation}
f = \frac{\Delta L_{\mathrm{IR}}}{\Delta L_{\mathrm{V}}} \\,
\end{equation}
where $\Delta L_{\mathrm{IR}}$ and $\Delta L_{\mathrm{V}}$ are the luminosity variability amplitudes of the IR and optical flares, respectively. The corresponding covering factors for WISE $W1$ and $W2$ are $0.7$ and $0.6$, respectively. These covering factors are similar to the results obtained by \cite{Stalevski2016} via the IR-to-bolometric luminosity ratio. 

We can compare the disk sizes with BLR and torus sizes. According to Equation~\ref{eq_tue}, the size of the $i$ band is $\simeq 6.8$ light days, which is smaller than the expected self-gravity radius \citep[$12$ light days;][]{Lobban2022} by a factor of $\sim 2$. The BLR size is expected to be $83.5_{-0.6}^{+0.5}$ light days according to the BLR size-luminosity relation (Section~\ref{sec:2.2}), which is $\sim 7$ times larger than the self-gravity radius. Following \cite{Lyu2019} (see their equation 9), we can estimate the dust sublimation radius from the luminosity \citep[also see][]{Barvainis1987}, which is $R_{\mathrm{sub}}\simeq 800$ light days. According to the WISE and optical light curves, the torus size in W1/W2 is $\sim 1000$ light days (Section~\ref{sec:3.1}); the WISE W1/W2 torus size is close to $R_{\mathrm{sub}}$ and is $\sim 10$ times larger than the BLR size, which is qualitatively consistent with the AGN unification model. The measured torus size can be reasonably treated as the outer boundary of the BLR. Hence, the BLR should be a very extended structure, roughly consistent with the dust-inflated accretion disk producing the BLR \citep[e.g.,][]{Baskin2018}.  

\section{Summary}\label{sec:5}
We used LCOGT to monitor a highly variable AGN, 6dFGS gJ022550.0-060145. Continuum lags between $u$, $g$, $r$, and $i$ bands have been detected in our target.  Here are our main conclusions:

\begin{itemize}
\item[1.] The rest-frame time delays of $g$, $r$, and $i$ bands with respect to the $u$ band are $5.2_{-3.2}^{+3.3}$ days, $2.3_{-2.2}^{+3.0}$ days, and $3.8_{-2.3}^{+2.5}$ days, respectively.
\item[2.] The black-hole mass from the single-epoch Lijiang 2.4m optical spectrum is $\log (M_{\mathrm{BH}}/M_{\odot})=8.02_{-0.02}^{+0.02}$. The ratio of the observed time delay to the SSD prediction is $2.6_{-1.3}^{+1.3}$. 

\item[3.]The measured damping timescale is larger than the $t_\mathrm{{damping}}$-$M_{\mathrm{BH}}$ relation of \cite{Burke2021} by a factor of $7$. 

\item[4.] Using the optical data from ZTF, PS1, CSS, and ATLAS, and the WISE IR light curves, we have measured time delays of IR $W1$ and $W2$ bands with optical emission, and the results are $9.6^{+2.9}_{-1.6}\times 10^2$ days and $1.18^{+0.13}_{-0.10}\times 10^3$ days, respectively. The two-time delays are close, which means that the dusty torus of 6dFGS gJ022550.0-060145 is relatively dense. 

\item[5.] 
The measured $W1$ and $W2$ time delays (with respect to the optical emission; Figure~\ref{fig:L-lag}) are larger than the IR lag-luminosity relation of \cite{Lyu2019}. The dust covering factors of the $W1$ and $W2$ emission regions are 0.7 and 0.6, respectively.
\end{itemize}

Future continuum reverberation mapping of AGNs with various $M_{\mathrm{BH}}$, $L_{\mathrm{AGN}}$, and other properties (e.g., the X-ray intensity; e.g., \citealt{Marcin2023}) can resolve the accretion-disk oversize problem and critically test the SMBH accretion physics. 

\section*{Acknowledgement}
We thank J. R. Trump and Gloria Fonseca Alvarez for helpful discussion in selecting targets. We thank Guowei Ren and Kaixing Lu for arranging the optical spectroscopic observation. We thank the referee for his/her helpful suggestions that greatly improved the manuscript. We thank Haikun Li for maintaining the computing resources. This work was supported by the National Key R\&D Program of China (No. 2023YFA1607903, No. 2023YFA1607904). D.Y.L. and M.Y.S. acknowledge support from the National Natural Science Foundation of China (NSFC-11973002), the Natural Science Foundation of Fujian Province of China (No. 2022J06002), and the China Manned Space Project grant (No. CMS-CSST-2021-A06 and CMS-CSST-2021-B11). J.Wang acknowledges support from the National Natural Science Foundation of China (NSFC-U1831205, NSFC-12033004, and NSFC-12221003), and the China Manned Space Project grants (No.\ CMS-CSST-2021-B02; No.\ CMS-CSST-2021-A06). J.Wu acknowledges support from the National Natural Science Foundation of China (NSFC-12273029 and NSFC-12221003), and the China Manned Space Project grants (No.\ CMS-CSST-2021-A05; No.\ CMS-CSST-2021-A06). Z.X.Z. acknowledges support from the National Natural Science Foundation of China (NSFC-12033006 and NSFC-12103041). Our LCOGT observation (CTAP2021-B0073) is kindly supported by China Telescope Access Program (TAP). 

Based on observations obtained with the Samuel Oschin Telescope 48-inch and the 60-inch Telescope at the Palomar
Observatory as part of the Zwicky Transient Facility project. ZTF is supported by the National Science Foundation under Grants
No. AST-1440341 and AST-2034437 and a collaboration including current partners Caltech, IPAC, the Weizmann Institute for
Science, the Oskar Klein Center at Stockholm University, the University of Maryland, Deutsches Elektronen-Synchrotron and
Humboldt University, the TANGO Consortium of Taiwan, the University of Wisconsin at Milwaukee, Trinity College Dublin,
Lawrence Livermore National Laboratories, IN2P3, University of Warwick, Ruhr University Bochum, Northwestern University and
former partners the University of Washington, Los Alamos National Laboratories, and Lawrence Berkeley National Laboratories.
Operations are conducted by COO, IPAC, and UW.

The Pan-STARRS1 Surveys (PS1) and the PS1 public science archive have been made possible through contributions by the Institute for Astronomy, the University of Hawaii, the Pan-STARRS Project Office, the Max-Planck Society and its participating institutes, the Max Planck Institute for Astronomy, Heidelberg and the Max Planck Institute for Extraterrestrial Physics, Garching, The Johns Hopkins University, Durham University, the University of Edinburgh, the Queen's University Belfast, the Harvard-Smithsonian Center for Astrophysics, the Las Cumbres Observatory Global Telescope Network Incorporated, the National Central University of Taiwan, the Space Telescope Science Institute, the National Aeronautics and Space Administration under Grant No. NNX08AR22G issued through the Planetary Science Division of the NASA Science Mission Directorate, the National Science Foundation Grant No. AST-1238877, the University of Maryland, Eotvos Lorand University (ELTE), the Los Alamos National Laboratory, and the Gordon and Betty Moore Foundation.

This work has made use of data from the Asteroid Terrestrial-impact Last Alert System (ATLAS) project. The Asteroid Terrestrial-impact Last Alert System (ATLAS) project is primarily funded to search for near earth asteroids through NASA grants NN12AR55G, 80NSSC18K0284, and 80NSSC18K1575; byproducts of the NEO search include images and catalogs from the survey area. This work was partially funded by Kepler/K2 grant J1944/80NSSC19K0112 and HST GO-15889, and STFC grants ST/T000198/1 and ST/S006109/1. The ATLAS science products have been made possible through the contributions of the University of Hawaii Institute for Astronomy, the Queen’s University Belfast, the Space Telescope Science Institute, the South African Astronomical Observatory, and The Millennium Institute of Astrophysics (MAS), Chile.

The CSS survey is funded by the National Aeronautics and Space
Administration under Grant No. NNG05GF22G issued through the Science Mission Directorate Near-Earth Objects Observations Program.  The CRTS survey is supported by the U.S. National Science Foundation under
grants AST-0909182.

Funding for the Sloan Digital Sky Survey IV has been provided by the Alfred P. Sloan Foundation, the U.S. Department of Energy Office of Science, and the Participating Institutions. SDSS acknowledges support and resources from the Center for High-Performance Computing at the University of Utah. The SDSS web site is \url{www.sdss.org}.

SDSS is managed by the Astrophysical Research Consortium for the Participating Institutions of the SDSS Collaboration including the Brazilian Participation Group, the Carnegie Institution for Science, Carnegie Mellon University, Center for Astrophysics | Harvard \& Smithsonian (CfA), the Chilean Participation Group, the French Participation Group, Instituto de Astrof\'{i}sica de Canarias, The Johns Hopkins University, Kavli Institute for the Physics and Mathematics of the Universe (IPMU) / University of Tokyo, the Korean Participation Group, Lawrence Berkeley National Laboratory, Leibniz Institut f\"{u}r Astrophysik Potsdam (AIP), Max-Planck-Institut f\"{u}r Astronomie (MPIA Heidelberg), Max-Planck-Institut für Astrophysik (MPA Garching), Max-Planck-Institut f\"{u}r Extraterrestrische Physik (MPE), National Astronomical Observatories of China, New Mexico State University, New York University, University of Notre Dame, Observatório Nacional / MCTI, The Ohio State University, Pennsylvania State University, Shanghai Astronomical Observatory, United Kingdom Participation Group, Universidad Nacional Aut\'{o}noma de M\'{e}xico, University of Arizona, University of Colorado Boulder, University of Oxford, University of Portsmouth, University of Utah, University of Virginia, University of Washington, University of Wisconsin, Vanderbilt University, and Yale University.

This publication makes use of data products from the Wide-field Infrared Survey Explorer, which is a joint project of the University of California, Los Angeles, and the Jet Propulsion Laboratory/California Institute of Technology, funded by the National Aeronautics and Space Administration.

This publication also makes use of data products from NEOWISE, which is a project of the Jet Propulsion Laboratory/California Institute of Technology, funded by the Planetary Science Division of the National Aeronautics and Space Administration.

This work makes use of observations from the Las Cumbres Observatory global telescope network. This paper is based on observations made with the 1.0 meter Sinistro instrument, including Tenerife (TFN), Dome B: fa11; Siding Spring (COJ), Dome B: fa19; Cerro Tololo (LSC), Dome C: fa03; Cerro Tololo (LSC), Dome A: fa15; McDonald (ELP), Dome B: fa07; McDonald (ELP), Dome A: fa05; Cerro Tololo (LSC), Dome B: fa04; Siding Spring (COJ), Dome A: fa12; Sutherland (CPT), Dome C: fa06; Sutherland (CPT), Dome A: fa14; Tenerife (TFN), Dome A: fa20.

\software{Astropy \citep{AstropyCollaboration2013}, \emph{PyCCF} \citep{Sun2018_ccf}, Matplotlib \citep{Hunter2007}, 
Numpy \& Scipy \citep{vanderWalt2011}}, \textit{emcee} \citep{Foreman-Mackey2013}, \textit{HEASoft} (\citealt{Nasa2014}), AutoPhOT \citep{Brennan2022}

\clearpage
\appendix

\section{AutoPhOT PERAMETERS}\label{sec:appendix}
We use AutoPHOT (version 1.0.8) to measure the PSF magnitudes of our target. The modified AutoPhOT parameters are listed in Table~\ref{table: autophot}. Other AutoPhOT parameters are set to the default values. The parameter meanings can be found from \url{https://autophot.readthedocs.io/en/latest/autophot_instructions.html}. We calibrated the $g$, $r$, $i$, and $u$ magnitudes against the custom catalog from the $16$th Data Release of the Sloan Digital Sky Surveys \citep{Ahumada2020}. The standard stars are selected according to the following criteria: 1. the angular distances between the stars and our target are less than 7 arcmin; 2. the stars are brighter than our target by $1\sim 3$ magnitudes; 3. the stars' variability amplitudes at $g$, $r$, and $i$ bands are smaller than 3\% according to their Pan-STARRS light curves \citep{Flewelling2020}. A total number of eleven standard stars are selected, and the resulting standard star catalog is used by AutoPHOT.

\begin{deluxetable}{ccc}
\tablecaption{AutoPhOT PERAMETERS}
\label{table: autophot}

\tablehead{\colhead{type} & \colhead{command} & \colhead{new\,value} } 
\startdata
photometry & force\_psf	& true\\
\hline
catalog	& use\_catalog	& custom \\
{} & matching\_source\_FWHM\_limit	& 3 \\
\hline
fitting	& use\_moffat & true \\
\hline
psf	& psf\_source\_no	& 20 \\
{} & plot\_PSF\_model\_residuals & true \\
{} & construction\_SNR & 40 \\
\hline
zeropoint & zp\_use\_median & true \\
{} & zp\_use\_fitted & true \\
{} & matching\_source\_SNR\_limit & 40
\enddata
\end{deluxetable}

\bibliography{ref}{}

\begin{thebibliography}{}
\expandafter\ifx\csname natexlab\endcsname\relax\def\natexlab#1{#1}\fi
\providecommand{\url}[1]{\href{#1}{#1}}
\providecommand{\dodoi}[1]{doi:~\href{http://doi.org/#1}{\nolinkurl{#1}}}
\providecommand{\doeprint}[1]{\href{http://ascl.net/#1}{\nolinkurl{http://ascl.net/#1}}}
\providecommand{\doarXiv}[1]{\href{https://arxiv.org/abs/#1}{\nolinkurl{https://arxiv.org/abs/#1}}}

\bibitem[{{Abramowicz} {et~al.}(1988){Abramowicz}, {Czerny}, {Lasota}, \& {Szuszkiewicz}}]{Abramowicz1988}
{Abramowicz}, M.~A., {Czerny}, B., {Lasota}, J.~P., \& {Szuszkiewicz}, E. 1988, \apj, 332, 646, \dodoi{10.1086/166683}

\bibitem[{{Ahumada} {et~al.}(2020){Ahumada}, {Allende Prieto}, {Almeida}, {Anders}, {Anderson}, {Andrews}, {Anguiano}, {Arcodia}, {Armengaud}, {Aubert}, {Avila}, {Avila-Reese}, {Badenes}, {Balland}, {Barger}, {Barrera-Ballesteros}, {Basu}, {Bautista}, {Beaton}, {Beers}, {Benavides}, {Bender}, {Bernardi}, {Bershady}, {Beutler}, {Bidin}, {Bird}, {Bizyaev}, {Blanc}, {Blanton}, {Boquien}, {Borissova}, {Bovy}, {Brandt}, {Brinkmann}, {Brownstein}, {Bundy}, {Bureau}, {Burgasser}, {Burtin}, {Cano-D{\'\i}az}, {Capasso}, {Cappellari}, {Carrera}, {Chabanier}, {Chaplin}, {Chapman}, {Cherinka}, {Chiappini}, {Doohyun Choi}, {Chojnowski}, {Chung}, {Clerc}, {Coffey}, {Comerford}, {Comparat}, {da Costa}, {Cousinou}, {Covey}, {Crane}, {Cunha}, {Ilha}, {Dai}, {Damsted}, {Darling}, {Davidson}, {Davies}, {Dawson}, {De}, {de la Macorra}, {De Lee}, {Queiroz}, {Deconto Machado}, {de la Torre}, {Dell'Agli}, {du Mas des Bourboux}, {Diamond-Stanic}, {Dillon}, {Donor}, {Drory}, {Duckworth}, {Dwelly}, {Ebelke}, {Eftekharzadeh}, {Davis Eigenbrot}, {Elsworth}, {Eracleous}, {Erfanianfar}, {Escoffier}, {Fan}, {Farr}, {Fern{\'a}ndez-Trincado}, {Feuillet}, {Finoguenov}, {Fofie}, {Fraser-McKelvie}, {Frinchaboy}, {Fromenteau}, {Fu}, {Galbany}, {Garcia}, {Garc{\'\i}a-Hern{\'a}ndez}, {Garma Oehmichen}, {Ge}, {Geimba Maia}, {Geisler}, {Gelfand}, {Goddy}, {Gonzalez-Perez}, {Grabowski}, {Green}, {Grier}, {Guo}, {Guy}, {Harding}, {Hasselquist}, {Hawken}, {Hayes}, {Hearty}, {Hekker}, {Hogg}, {Holtzman}, {Horta}, {Hou}, {Hsieh}, {Huber}, {Hunt}, {Ider Chitham}, {Imig}, {Jaber}, {Jimenez Angel}, {Johnson}, {Jones}, {J{\"o}nsson}, {Jullo}, {Kim}, {Kinemuchi}, {Kirkpatrick}, {Kite}, {Klaene}, {Kneib}, {Kollmeier}, {Kong}, {Kounkel}, {Krishnarao}, {Lacerna}, {Lan}, {Lane}, {Law}, {Le Goff}, {Leung}, {Lewis}, {Li}, {Lian}, {Lin}, {Long}, {Longa-Pe{\~n}a}, {Lundgren}, {Lyke}, {Mackereth}, {MacLeod}, {Majewski}, {Manchado}, {Maraston}, {Martini}, {Masseron}, {Masters}, {Mathur}, {McDermid}, {Merloni}, {Merrifield}, {M{\'e}sz{\'a}ros}, {Miglio}, {Minniti}, {Minsley}, {Miyaji}, {Mohammad}, {Mosser}, {Mueller}, {Muna}, {Mu{\~n}oz-Guti{\'e}rrez}, {Myers}, {Nadathur}, {Nair}, {Nandra}, {Correa do Nascimento}, {Nevin}, {Newman}, {Nidever}, {Nitschelm}, {Noterdaeme}, {O'Connell}, {Olmstead}, {Oravetz}, {Oravetz}, {Osorio}, {Pace}, {Padilla}, {Palanque-Delabrouille}, {Palicio}, {Pan}, {Pan}, {Parker}, {Paviot}, {Peirani}, {Ram{\'r}ez}, {Penny}, {Percival}, {Perez-Fournon}, {P{\'e}rez-R{\`a}fols}, {Petitjean}, {Pieri}, {Pinsonneault}, {Poovelil}, {Povick}, {Prakash}, {Price-Whelan}, {Raddick}, {Raichoor}, {Ray}, {Rembold}, {Rezaie}, {Riffel}, {Riffel}, {Rix}, {Robin}, {Roman-Lopes}, {Rom{\'a}n-Z{\'u}{\~n}iga}, {Rose}, {Ross}, {Rossi}, {Rowlands}, {Rubin}, {Salvato}, {S{\'a}nchez}, {S{\'a}nchez-Menguiano}, {S{\'a}nchez-Gallego}, {Sayres}, {Schaefer}, {Schiavon}, {Schimoia}, {Schlafly}, {Schlegel}, {Schneider}, {Schultheis}, {Schwope}, {Seo}, {Serenelli}, {Shafieloo}, {Shamsi}, {Shao}, {Shen}, {Shetrone}, {Shirley}, {Silva Aguirre}, {Simon}, {Skrutskie}, {Slosar}, {Smethurst}, {Sobeck}, {Sodi}, {Souto}, {Stark}, {Stassun}, {Steinmetz}, {Stello}, {Stermer}, {Storchi-Bergmann}, {Streblyanska}, {Stringfellow}, {Stutz}, {Su{\'a}rez}, {Sun}, {Taghizadeh-Popp}, {Talbot}, {Tayar}, {Thakar}, {Theriault}, {Thomas}, {Thomas}, {Tinker}, {Tojeiro}, {Toledo}, {Tremonti}, {Troup}, {Tuttle}, {Unda-Sanzana}, {Valentini}, {Vargas-Gonz{\'a}lez}, {Vargas-Maga{\~n}a}, {V{\'a}zquez-Mata}, {Vivek}, {Wake}, {Wang}, {Weaver}, {Weijmans}, {Wild}, {Wilson}, {Wilson}, {Wolthuis}, {Wood-Vasey}, {Yan}, {Yang}, {Y{\`e}che}, {Zamora}, {Zarrouk}, {Zasowski}, {Zhang}, {Zhao}, {Zhao}, {Zheng}, {Zheng}, {Zhu}, \& {Zou}}]{Ahumada2020}
{Ahumada}, R., {Allende Prieto}, C., {Almeida}, A., {et~al.} 2020, \apjs, 249, 3, \dodoi{10.3847/1538-4365/ab929e}

\bibitem[{{Astropy Collaboration} {et~al.}(2013){Astropy Collaboration}, {Robitaille}, {Tollerud}, {Greenfield}, {Droettboom}, {Bray}, {Aldcroft}, {Davis}, {Ginsburg}, {Price-Whelan}, {Kerzendorf}, {Conley}, {Crighton}, {Barbary}, {Muna}, {Ferguson}, {Grollier}, {Parikh}, {Nair}, {Unther}, {Deil}, {Woillez}, {Conseil}, {Kramer}, {Turner}, {Singer}, {Fox}, {Weaver}, {Zabalza}, {Edwards}, {Azalee Bostroem}, {Burke}, {Casey}, {Crawford}, {Dencheva}, {Ely}, {Jenness}, {Labrie}, {Lim}, {Pierfederici}, {Pontzen}, {Ptak}, {Refsdal}, {Servillat}, \& {Streicher}}]{AstropyCollaboration2013}
{Astropy Collaboration}, {Robitaille}, T.~P., {Tollerud}, E.~J., {et~al.} 2013, \aap, 558, A33, \dodoi{10.1051/0004-6361/201322068}

\bibitem[{{Barvainis}(1987)}]{Barvainis1987}
{Barvainis}, R. 1987, \apj, 320, 537, \dodoi{10.1086/165571}

\bibitem[{{Barvainis}(1992)}]{Barvainis1992}
---. 1992, \apj, 400, 502, \dodoi{10.1086/172012}

\bibitem[{{Baskin} \& {Laor}(2018)}]{Baskin2018}
{Baskin}, A., \& {Laor}, A. 2018, \mnras, 474, 1970, \dodoi{10.1093/mnras/stx2850}

\bibitem[{{Bentz} {et~al.}(2009){Bentz}, {Peterson}, {Netzer}, {Pogge}, \& {Vestergaard}}]{Bentz2009}
{Bentz}, M.~C., {Peterson}, B.~M., {Netzer}, H., {Pogge}, R.~W., \& {Vestergaard}, M. 2009, \apj, 697, 160, \dodoi{10.1088/0004-637X/697/1/160}

\bibitem[{{Brennan} \& {Fraser}(2022)}]{Brennan2022}
{Brennan}, S.~J., \& {Fraser}, M. 2022, \aap, 667, A62, \dodoi{10.1051/0004-6361/202243067}

\bibitem[{{Burbidge}(1967)}]{Burbidge1967}
{Burbidge}, E.~M. 1967, \araa, 5, 399, \dodoi{10.1146/annurev.aa.05.090167.002151}

\bibitem[{{Burke} {et~al.}(2021){Burke}, {Shen}, {Blaes}, {Gammie}, {Horne}, {Jiang}, {Liu}, {McHardy}, {Morgan}, {Scaringi}, \& {Yang}}]{Burke2021}
{Burke}, C.~J., {Shen}, Y., {Blaes}, O., {et~al.} 2021, Science, 373, 789, \dodoi{10.1126/science.abg9933}

\bibitem[{{Cackett} {et~al.}(2021){Cackett}, {Bentz}, \& {Kara}}]{Cackett2021}
{Cackett}, E.~M., {Bentz}, M.~C., \& {Kara}, E. 2021, iScience, 24, 102557, \dodoi{10.1016/j.isci.2021.102557}

\bibitem[{{Cackett} {et~al.}(2018){Cackett}, {Chiang}, {McHardy}, {Edelson}, {Goad}, {Horne}, \& {Korista}}]{Cackett2018}
{Cackett}, E.~M., {Chiang}, C.-Y., {McHardy}, I., {et~al.} 2018, \apj, 857, 53, \dodoi{10.3847/1538-4357/aab4f7}

\bibitem[{{Cackett} {et~al.}(2007){Cackett}, {Horne}, \& {Winkler}}]{Cackett2007}
{Cackett}, E.~M., {Horne}, K., \& {Winkler}, H. 2007, \mnras, 380, 669, \dodoi{10.1111/j.1365-2966.2007.12098.x}

\bibitem[{{Cackett} {et~al.}(2022){Cackett}, {Zoghbi}, \& {Ulrich}}]{Cackett2022}
{Cackett}, E.~M., {Zoghbi}, A., \& {Ulrich}, O. 2022, \apj, 925, 29, \dodoi{10.3847/1538-4357/ac3913}

\bibitem[{{Cackett} {et~al.}(2023){Cackett}, {Gelbord}, {Barth}, {De Rosa}, {Edelson}, {Goad}, {Homayouni}, {Horne}, {Kara}, {Kriss}, {Korista}, {Landt}, {Plesha}, {Arav}, {Bentz}, {Boizelle}, {Dalla Bont{\`a}}, {Dehghanian}, {Donnan}, {Du}, {Ferland}, {Fian}, {Filippenko}, {Gonz{\'a}lez Buitrago}, {Grier}, {Hall}, {Hu}, {Ili{\'c}}, {Kaastra}, {Kaspi}, {Kochanek}, {Kova{\v{c}}evi{\'c}}, {Kynoch}, {Li}, {McLane}, {Mehdipour}, {Miller}, {Montano}, {Netzer}, {Panagiotou}, {Partington}, {{\v{C}}. Popovi{\'c}}, {Proga}, {Rogantini}, {Sanmartim}, {Siebert}, {Storchi-Bergmann}, {Vestergaard}, {Wang}, {Waters}, \& {Zaidouni}}]{Cackett2023}
{Cackett}, E.~M., {Gelbord}, J., {Barth}, A.~J., {et~al.} 2023, \apj, 958, 195, \dodoi{10.3847/1538-4357/acfdac}

\bibitem[{{Clavel} {et~al.}(1991){Clavel}, {Reichert}, {Alloin}, {Crenshaw}, {Kriss}, {Krolik}, {Malkan}, {Netzer}, {Peterson}, {Wamsteker}, {Altamore}, {Baribaud}, {Barr}, {Beck}, {Binette}, {Bromage}, {Brosch}, {Diaz}, {Filippenko}, {Fricke}, {Gaskell}, {Giommi}, {Glass}, {Gondhalekar}, {Hackney}, {Halpern}, {Hutter}, {Joersaeter}, {Kinney}, {Kollatschny}, {Koratkar}, {Korista}, {Laor}, {Lasota}, {Leibowitz}, {Maoz}, {Martin}, {Mazeh}, {Meurs}, {Nair}, {O'Brien}, {Pelat}, {Perez}, {Perola}, {Ptak}, {Rodriguez-Pascual}, {Rosenblatt}, {Sadun}, {Santos-Lleo}, {Shaw}, {Smith}, {Stirpe}, {Stoner}, {Sun}, {Ulrich}, {van Groningen}, \& {Zheng}}]{Clavel1991}
{Clavel}, J., {Reichert}, G.~A., {Alloin}, D., {et~al.} 1991, \apj, 366, 64, \dodoi{10.1086/169540}

\bibitem[{{Dexter} {et~al.}(2019){Dexter}, {Xin}, {Shen}, {Grier}, {Liu}, {Gezari}, {McGreer}, {Brandt}, {Hall}, {Horne}, {Simm}, {Merloni}, {Green}, {Vivek}, {Trump}, {Homayouni}, {Peterson}, {Schneider}, {Kinemuchi}, {Pan}, \& {Bizyaev}}]{Dexter2019}
{Dexter}, J., {Xin}, S., {Shen}, Y., {et~al.} 2019, \apj, 885, 44, \dodoi{10.3847/1538-4357/ab4354}

\bibitem[{{Drake} {et~al.}(2009){Drake}, {Djorgovski}, {Mahabal}, {Beshore}, {Larson}, {Graham}, {Williams}, {Christensen}, {Catelan}, {Boattini}, {Gibbs}, {Hill}, \& {Kowalski}}]{Drake2009}
{Drake}, A.~J., {Djorgovski}, S.~G., {Mahabal}, A., {et~al.} 2009, \apj, 696, 870, \dodoi{10.1088/0004-637X/696/1/870}

\bibitem[{Du {et~al.}(2016)Du, Lu, Zhang, Huang, Wang, Hu, Qiu, Li, Fan, Fang, {et~al.}}]{Du2016}
Du, P., Lu, K.-X., Zhang, Z.-X., {et~al.} 2016, The Astrophysical Journal, 825, 126

\bibitem[{Du {et~al.}(2018)Du, Zhang, Wang, Huang, Zhang, Lu, Hu, Li, Bai, Bian, {et~al.}}]{Du2018}
Du, P., Zhang, Z.-X., Wang, K., {et~al.} 2018, The Astrophysical Journal, 856, 6

\bibitem[{{Fausnaugh} {et~al.}(2016){Fausnaugh}, {Denney}, {Barth}, {Bentz}, {Bottorff}, {Carini}, {Croxall}, {De Rosa}, {Goad}, {Horne}, {Joner}, {Kaspi}, {Kim}, {Klimanov}, {Kochanek}, {Leonard}, {Netzer}, {Peterson}, {Schn{\"u}lle}, {Sergeev}, {Vestergaard}, {Zheng}, {Zu}, {Anderson}, {Ar{\'e}valo}, {Bazhaw}, {Borman}, {Boroson}, {Brandt}, {Breeveld}, {Brewer}, {Cackett}, {Crenshaw}, {Dalla Bont{\`a}}, {De Lorenzo-C{\'a}ceres}, {Dietrich}, {Edelson}, {Efimova}, {Ely}, {Evans}, {Filippenko}, {Flatland}, {Gehrels}, {Geier}, {Gelbord}, {Gonzalez}, {Gorjian}, {Grier}, {Grupe}, {Hall}, {Hicks}, {Horenstein}, {Hutchison}, {Im}, {Jensen}, {Jones}, {Kaastra}, {Kelly}, {Kennea}, {Kim}, {Korista}, {Kriss}, {Lee}, {Lira}, {MacInnis}, {Manne-Nicholas}, {Mathur}, {McHardy}, {Montouri}, {Musso}, {Nazarov}, {Norris}, {Nousek}, {Okhmat}, {Pancoast}, {Papadakis}, {Parks}, {Pei}, {Pogge}, {Pott}, {Rafter}, {Rix}, {Saylor}, {Schimoia}, {Siegel}, {Spencer}, {Starkey}, {Sung}, {Teems}, {Treu}, {Turner}, {Uttley}, {Villforth}, {Weiss}, {Woo}, {Yan}, \& {Young}}]{Fausnaugh2016}
{Fausnaugh}, M.~M., {Denney}, K.~D., {Barth}, A.~J., {et~al.} 2016, \apj, 821, 56, \dodoi{10.3847/0004-637X/821/1/56}

\bibitem[{{Fian} {et~al.}(2023){Fian}, {Chelouche}, {Kaspi}, {Sobrino Figaredo}, {Lewis}, \& {Catalan}}]{Fian2023}
{Fian}, C., {Chelouche}, D., {Kaspi}, S., {et~al.} 2023, \aap, 672, A132, \dodoi{10.1051/0004-6361/202244905}

\bibitem[{{Flewelling} {et~al.}(2020){Flewelling}, {Magnier}, {Chambers}, {Heasley}, {Holmberg}, {Huber}, {Sweeney}, {Waters}, {Calamida}, {Casertano}, {Chen}, {Farrow}, {Hasinger}, {Henderson}, {Long}, {Metcalfe}, {Narayan}, {Nieto-Santisteban}, {Norberg}, {Rest}, {Saglia}, {Szalay}, {Thakar}, {Tonry}, {Valenti}, {Werner}, {White}, {Denneau}, {Draper}, {Hodapp}, {Jedicke}, {Kaiser}, {Kudritzki}, {Price}, {Wainscoat}, {Chastel}, {McLean}, {Postman}, \& {Shiao}}]{Flewelling2020}
{Flewelling}, H.~A., {Magnier}, E.~A., {Chambers}, K.~C., {et~al.} 2020, \apjs, 251, 7, \dodoi{10.3847/1538-4365/abb82d}

\bibitem[{{Fonseca Alvarez} {et~al.}(2020){Fonseca Alvarez}, {Trump}, {Homayouni}, {Grier}, {Shen}, {Horne}, {Li}, {Brandt}, {Ho}, {Peterson}, \& {Schneider}}]{Fonseca2020}
{Fonseca Alvarez}, G., {Trump}, J.~R., {Homayouni}, Y., {et~al.} 2020, \apj, 899, 73, \dodoi{10.3847/1538-4357/aba001}

\bibitem[{{Foreman-Mackey} {et~al.}(2013){Foreman-Mackey}, {Hogg}, {Lang}, \& {Goodman}}]{Foreman-Mackey2013}
{Foreman-Mackey}, D., {Hogg}, D.~W., {Lang}, D., \& {Goodman}, J. 2013, \pasp, 125, 306, \dodoi{10.1086/670067}

\bibitem[{{Gardner} \& {Done}(2017)}]{Gardner2017}
{Gardner}, E., \& {Done}, C. 2017, \mnras, 470, 3591, \dodoi{10.1093/mnras/stx946}

\bibitem[{{Gaskell}(2017)}]{Gaskell2017}
{Gaskell}, C.~M. 2017, \mnras, 467, 226, \dodoi{10.1093/mnras/stx094}

\bibitem[{{Guo} {et~al.}(2022{\natexlab{a}}){Guo}, {Barth}, \& {Wang}}]{GuoHengxiao2022}
{Guo}, H., {Barth}, A.~J., \& {Wang}, S. 2022{\natexlab{a}}, \apj, 940, 20, \dodoi{10.3847/1538-4357/ac96ec}

\bibitem[{{Guo} {et~al.}(2022{\natexlab{b}}){Guo}, {Li}, {Zhang}, {Ho}, \& {Wang}}]{Guo_Wei2022}
{Guo}, W.-J., {Li}, Y.-R., {Zhang}, Z.-X., {Ho}, L.~C., \& {Wang}, J.-M. 2022{\natexlab{b}}, \apj, 929, 19, \dodoi{10.3847/1538-4357/ac4e84}

\bibitem[{{Hall} {et~al.}(2018){Hall}, {Sarrouh}, \& {Horne}}]{Hall2018}
{Hall}, P.~B., {Sarrouh}, G.~T., \& {Horne}, K. 2018, \apj, 854, 93, \dodoi{10.3847/1538-4357/aaa768}

\bibitem[{{Homayouni} {et~al.}(2019){Homayouni}, {Trump}, {Grier}, {Shen}, {Starkey}, {Brandt}, {Fonseca Alvarez}, {Hall}, {Horne}, {Kinemuchi}, {I-Hsiu Li}, {McGreer}, {Sun}, {Ho}, \& {Schneider}}]{Homayouni2019}
{Homayouni}, Y., {Trump}, J.~R., {Grier}, C.~J., {et~al.} 2019, \apj, 880, 126, \dodoi{10.3847/1538-4357/ab2638}

\bibitem[{{Hu} {et~al.}(2023){Hu}, {Cai}, \& {Wang}}]{Hu2023}
{Hu}, X.-F., {Cai}, Z.-Y., \& {Wang}, J.-X. 2023, arXiv e-prints, arXiv:2310.16223, \dodoi{10.48550/arXiv.2310.16223}

\bibitem[{{Hunter}(2007)}]{Hunter2007}
{Hunter}, J.~D. 2007, Computing in Science and Engineering, 9, 90, \dodoi{10.1109/MCSE.2007.55}

\bibitem[{{Jester} {et~al.}(2005){Jester}, {Schneider}, {Richards}, {Green}, {Schmidt}, {Hall}, {Strauss}, {Vanden Berk}, {Stoughton}, {Gunn}, {Brinkmann}, {Kent}, {Smith}, {Tucker}, \& {Yanny}}]{Jester2005}
{Jester}, S., {Schneider}, D.~P., {Richards}, G.~T., {et~al.} 2005, \aj, 130, 873, \dodoi{10.1086/432466}

\bibitem[{{Kammoun} {et~al.}(2021){Kammoun}, {Dov{\v{c}}iak}, {Papadakis}, {Caballero-Garc{\'\i}a}, \& {Karas}}]{Kammoun2021}
{Kammoun}, E.~S., {Dov{\v{c}}iak}, M., {Papadakis}, I.~E., {Caballero-Garc{\'\i}a}, M.~D., \& {Karas}, V. 2021, \apj, 907, 20, \dodoi{10.3847/1538-4357/abcb93}

\bibitem[{{Kelly} {et~al.}(2009){Kelly}, {Bechtold}, \& {Siemiginowska}}]{Kelly2009}
{Kelly}, B.~C., {Bechtold}, J., \& {Siemiginowska}, A. 2009, \apj, 698, 895, \dodoi{10.1088/0004-637X/698/1/895}

\bibitem[{{Koshida} {et~al.}(2014){Koshida}, {Minezaki}, {Yoshii}, {Kobayashi}, {Sakata}, {Sugawara}, {Enya}, {Suganuma}, {Tomita}, {Aoki}, \& {Peterson}}]{Koshida2014}
{Koshida}, S., {Minezaki}, T., {Yoshii}, Y., {et~al.} 2014, \apj, 788, 159, \dodoi{10.1088/0004-637X/788/2/159}

\bibitem[{{Koz{\l}owski}(2017)}]{Kozlowski2017}
{Koz{\l}owski}, S. 2017, \aap, 597, A128, \dodoi{10.1051/0004-6361/201629890}

\bibitem[{{Krolik} {et~al.}(1991){Krolik}, {Horne}, {Kallman}, {Malkan}, {Edelson}, \& {Kriss}}]{Krolik1991}
{Krolik}, J.~H., {Horne}, K., {Kallman}, T.~R., {et~al.} 1991, \apj, 371, 541, \dodoi{10.1086/169918}

\bibitem[{{Li} {et~al.}(2021){Li}, {Sun}, {Xu}, {Brandt}, {Trump}, {Yu}, {Wang}, {Xue}, {Cai}, {Gu}, {Homayouni}, {Liu}, {Wang}, {Zhang}, \& {Li}}]{Li2021}
{Li}, T., {Sun}, M., {Xu}, X., {et~al.} 2021, \apjl, 912, L29, \dodoi{10.3847/2041-8213/abf9aa}

\bibitem[{{Lobban} \& {King}(2022)}]{Lobban2022}
{Lobban}, A., \& {King}, A. 2022, \mnras, 511, 1992, \dodoi{10.1093/mnras/stac155}

\bibitem[{{Lu} {et~al.}(2019){Lu}, {Bai}, {Zhang}, {Du}, {Hu}, {Kim}, {Wang}, {Ho}, {Li}, {Bian}, {Yuan}, {Xiao}, {Feng}, {Wang}, {Xu}, {Ding}, {Yu}, {Xin}, {Ye}, {Wang}, {Lun}, {Zhang}, {Zhang}, {Ji}, {Fan}, \& {Chang}}]{Lu2019}
{Lu}, K.-X., {Bai}, J.-M., {Zhang}, Z.-X., {et~al.} 2019, \apj, 887, 135, \dodoi{10.3847/1538-4357/ab5790}

\bibitem[{{Lu} {et~al.}(2021){Lu}, {Zhang}, {Huang}, {Ren}, {Xu}, {Feng}, {Xin}, {Ding}, {Yu}, \& {Bai}}]{lu2021}
{Lu}, K.-X., {Zhang}, Z.-X., {Huang}, Y.-K., {et~al.} 2021, Research in Astronomy and Astrophysics, 21, 183, \dodoi{10.1088/1674-4527/21/7/183}

\bibitem[{{Lyu} {et~al.}(2019){Lyu}, {Rieke}, \& {Smith}}]{Lyu2019}
{Lyu}, J., {Rieke}, G.~H., \& {Smith}, P.~S. 2019, \apj, 886, 33, \dodoi{10.3847/1538-4357/ab481d}

\bibitem[{{Mainzer} {et~al.}(2011){Mainzer}, {Bauer}, {Grav}, {Masiero}, {Cutri}, {Dailey}, {Eisenhardt}, {McMillan}, {Wright}, {Walker}, {Jedicke}, {Spahr}, {Tholen}, {Alles}, {Beck}, {Brandenburg}, {Conrow}, {Evans}, {Fowler}, {Jarrett}, {Marsh}, {Masci}, {McCallon}, {Wheelock}, {Wittman}, {Wyatt}, {DeBaun}, {Elliott}, {Elsbury}, {Gautier}, {Gomillion}, {Leisawitz}, {Maleszewski}, {Micheli}, \& {Wilkins}}]{Mainzer2011}
{Mainzer}, A., {Bauer}, J., {Grav}, T., {et~al.} 2011, \apj, 731, 53, \dodoi{10.1088/0004-637X/731/1/53}

\bibitem[{{Marculewicz} {et~al.}(2023){Marculewicz}, {Sun}, {Wu}, \& {Zhang}}]{Marcin2023}
{Marculewicz}, M., {Sun}, M., {Wu}, J., \& {Zhang}, Z. 2023, arXiv e-prints, arXiv:2308.11310, \dodoi{10.48550/arXiv.2308.11310}

\bibitem[{Masci {et~al.}(2018)Masci, Laher, Rusholme, Shupe, Groom, Surace, Jackson, Monkewitz, Beck, Flynn, Terek, Landry, Hacopians, Desai, Howell, Brooke, Imel, Wachter, Ye, Lin, Cenko, Cunningham, Rebbapragada, Bue, Miller, Mahabal, Bellm, Patterson, Jurić, Golkhou, Ofek, Walters, Graham, Kasliwal, Dekany, Kupfer, Burdge, Cannella, Barlow, Sistine, Giomi, Fremling, Blagorodnova, Levitan, Riddle, Smith, Helou, Prince, \& Kulkarni}]{Masci2019}
Masci, F.~J., Laher, R.~R., Rusholme, B., {et~al.} 2018, Publications of the Astronomical Society of the Pacific, 131, 018003, \dodoi{10.1088/1538-3873/aae8ac}

\bibitem[{{McHardy} {et~al.}(2018){McHardy}, {Connolly}, {Horne}, {Cackett}, {Gelbord}, {Peterson}, {Pahari}, {Gehrels}, {Goad}, {Lira}, {Arevalo}, {Baldi}, {Brandt}, {Breedt}, {Chand}, {Dewangan}, {Done}, {Elvis}, {Emmanoulopoulos}, {Fausnaugh}, {Kaspi}, {Kochanek}, {Korista}, {Papadakis}, {Rao}, {Uttley}, {Vestergaard}, \& {Ward}}]{McHardy2018}
{McHardy}, I.~M., {Connolly}, S.~D., {Horne}, K., {et~al.} 2018, \mnras, 480, 2881, \dodoi{10.1093/mnras/sty1983}

\bibitem[{{Monroe} {et~al.}(2016){Monroe}, {Prochaska}, {Tejos}, {Worseck}, {Hennawi}, {Schmidt}, {Tumlinson}, \& {Shen}}]{Monroe2016}
{Monroe}, T.~R., {Prochaska}, J.~X., {Tejos}, N., {et~al.} 2016, \aj, 152, 25, \dodoi{10.3847/0004-6256/152/1/25}

\bibitem[{{Narayan} \& {Yi}(1995)}]{Narayan1995}
{Narayan}, R., \& {Yi}, I. 1995, \apj, 452, 710, \dodoi{10.1086/176343}

\bibitem[{{Nasa High Energy Astrophysics Science Archive Research Center (Heasarc)}(2014)}]{Nasa2014}
{Nasa High Energy Astrophysics Science Archive Research Center (Heasarc)}. 2014, {HEAsoft: Unified Release of FTOOLS and XANADU}, Astrophysics Source Code Library, record ascl:1408.004.
\newblock \doeprint{1408.004}

\bibitem[{{Netzer}(2015)}]{Netzer2015}
{Netzer}, H. 2015, \araa, 53, 365, \dodoi{10.1146/annurev-astro-082214-122302}

\bibitem[{{Neustadt} \& {Kochanek}(2022)}]{Neustadt2022}
{Neustadt}, J.~M.~M., \& {Kochanek}, C.~S. 2022, \mnras, 513, 1046, \dodoi{10.1093/mnras/stac888}

\bibitem[{{Page} \& {Thorne}(1974)}]{Page1974}
{Page}, D.~N., \& {Thorne}, K.~S. 1974, \apj, 191, 499, \dodoi{10.1086/152990}

\bibitem[{{Runnoe} {et~al.}(2012){Runnoe}, {Brotherton}, \& {Shang}}]{Runnoe2012}
{Runnoe}, J.~C., {Brotherton}, M.~S., \& {Shang}, Z. 2012, \mnras, 422, 478, \dodoi{10.1111/j.1365-2966.2012.20620.x}

\bibitem[{{Schmidt} {et~al.}(2012){Schmidt}, {Rix}, {Shields}, {Knecht}, {Hogg}, {Maoz}, \& {Bovy}}]{Schmidt2012}
{Schmidt}, K.~B., {Rix}, H.-W., {Shields}, J.~C., {et~al.} 2012, \apj, 744, 147, \dodoi{10.1088/0004-637X/744/2/147}

\bibitem[{{Secunda} {et~al.}(2023){Secunda}, {Jiang}, \& {Greene}}]{Secunda2023}
{Secunda}, A., {Jiang}, Y.-F., \& {Greene}, J.~E. 2023, arXiv e-prints, arXiv:2311.10820, \dodoi{10.48550/arXiv.2311.10820}

\bibitem[{{Shakura} \& {Sunyaev}(1973)}]{Shakura1973}
{Shakura}, N.~I., \& {Sunyaev}, R.~A. 1973, \aap, 24, 337

\bibitem[{{Sharp} {et~al.}(2023){Sharp}, {Homayouni}, {Trump}, {Anderson}, {Assef}, {Brandt}, {Davis}, {Fries}, {Grier}, {Hall}, {Horne}, {Koekemoer}, {Mart{\'\i}nez-Aldama}, {Menezes}, {Pena}, {Ricci}, {Schneider}, {Shen}, \& {Trakhtenbrot}}]{Sharp2023}
{Sharp}, H.~W., {Homayouni}, Y., {Trump}, J.~R., {et~al.} 2023, arXiv e-prints, arXiv:2309.02499, \dodoi{10.48550/arXiv.2309.02499}

\bibitem[{{Shen} {et~al.}(2011){Shen}, {Richards}, {Strauss}, {Hall}, {Schneider}, {Snedden}, {Bizyaev}, {Brewington}, {Malanushenko}, {Malanushenko}, {Oravetz}, {Pan}, \& {Simmons}}]{Shen2011}
{Shen}, Y., {Richards}, G.~T., {Strauss}, M.~A., {et~al.} 2011, \apjs, 194, 45, \dodoi{10.1088/0067-0049/194/2/45}

\bibitem[{{Shields}(1978)}]{Shields1978}
{Shields}, G.~A. 1978, \nat, 272, 706, \dodoi{10.1038/272706a0}

\bibitem[{{Smith} {et~al.}(2020){Smith}, {Smartt}, {Young}, {Tonry}, {Denneau}, {Flewelling}, {Heinze}, {Weiland}, {Stalder}, {Rest}, {Stubbs}, {Anderson}, {Chen}, {Clark}, {Do}, {F{\"o}rster}, {Fulton}, {Gillanders}, {McBrien}, {O'Neill}, {Srivastav}, \& {Wright}}]{Smith2020}
{Smith}, K.~W., {Smartt}, S.~J., {Young}, D.~R., {et~al.} 2020, \pasp, 132, 085002, \dodoi{10.1088/1538-3873/ab936e}

\bibitem[{{Stalevski} {et~al.}(2016){Stalevski}, {Ricci}, {Ueda}, {Lira}, {Fritz}, \& {Baes}}]{Stalevski2016}
{Stalevski}, M., {Ricci}, C., {Ueda}, Y., {et~al.} 2016, \mnras, 458, 2288, \dodoi{10.1093/mnras/stw444}

\bibitem[{{Starkey} {et~al.}(2023){Starkey}, {Huang}, {Horne}, \& {Lin}}]{Starkey2023}
{Starkey}, D.~A., {Huang}, J., {Horne}, K., \& {Lin}, D. N.~C. 2023, \mnras, 519, 2754, \dodoi{10.1093/mnras/stac3579}

\bibitem[{{Steffen} {et~al.}(2006){Steffen}, {Strateva}, {Brandt}, {Alexander}, {Koekemoer}, {Lehmer}, {Schneider}, \& {Vignali}}]{Steffen2006}
{Steffen}, A.~T., {Strateva}, I., {Brandt}, W.~N., {et~al.} 2006, \aj, 131, 2826, \dodoi{10.1086/503627}

\bibitem[{{Sun} {et~al.}(2018{\natexlab{a}}){Sun}, {Grier}, \& {Peterson}}]{Sun2018_ccf}
{Sun}, M., {Grier}, C.~J., \& {Peterson}, B.~M. 2018{\natexlab{a}}, {PyCCF: Python Cross Correlation Function for reverberation mapping studies}, Astrophysics Source Code Library, record ascl:1805.032.
\newblock \doeprint{1805.032}

\bibitem[{{Sun} {et~al.}(2018{\natexlab{b}}){Sun}, {Xue}, {Cai}, \& {Guo}}]{Sun2018}
{Sun}, M., {Xue}, Y., {Cai}, Z., \& {Guo}, H. 2018{\natexlab{b}}, \apj, 857, 86, \dodoi{10.3847/1538-4357/aab786}

\bibitem[{{Sun} {et~al.}(2019){Sun}, {Xue}, {Trump}, \& {Gu}}]{Sun2019}
{Sun}, M., {Xue}, Y., {Trump}, J.~R., \& {Gu}, W.-M. 2019, \mnras, 482, 2788, \dodoi{10.1093/mnras/sty2885}

\bibitem[{{Sun} {et~al.}(2015){Sun}, {Trump}, {Brandt}, {Luo}, {Alexander}, {Jahnke}, {Rosario}, {Wang}, \& {Xue}}]{Sun2015}
{Sun}, M., {Trump}, J.~R., {Brandt}, W.~N., {et~al.} 2015, \apj, 802, 14, \dodoi{10.1088/0004-637X/802/1/14}

\bibitem[{{Sun} {et~al.}(2020){Sun}, {Xue}, {Brandt}, {Gu}, {Trump}, {Cai}, {He}, {Lin}, {Liu}, \& {Wang}}]{Sun2020}
{Sun}, M., {Xue}, Y., {Brandt}, W.~N., {et~al.} 2020, \apj, 891, 178, \dodoi{10.3847/1538-4357/ab789e}

\bibitem[{{Sun} {et~al.}(2014){Sun}, {Wang}, {Chen}, \& {Zheng}}]{Sun2014}
{Sun}, Y.-H., {Wang}, J.-X., {Chen}, X.-Y., \& {Zheng}, Z.-Y. 2014, \apj, 792, 54, \dodoi{10.1088/0004-637X/792/1/54}

\bibitem[{{Tie} \& {Kochanek}(2018)}]{Tie2018}
{Tie}, S.~S., \& {Kochanek}, C.~S. 2018, \mnras, 473, 80, \dodoi{10.1093/mnras/stx2348}

\bibitem[{{Tonry} {et~al.}(2018){Tonry}, {Denneau}, {Heinze}, {Stalder}, {Smith}, {Smartt}, {Stubbs}, {Weiland}, \& {Rest}}]{Tonry2018}
{Tonry}, J.~L., {Denneau}, L., {Heinze}, A.~N., {et~al.} 2018, \pasp, 130, 064505, \dodoi{10.1088/1538-3873/aabadf}

\bibitem[{{van der Walt} {et~al.}(2011){van der Walt}, {Colbert}, \& {Varoquaux}}]{vanderWalt2011}
{van der Walt}, S., {Colbert}, S.~C., \& {Varoquaux}, G. 2011, Computing in Science and Engineering, 13, 22, \dodoi{10.1109/MCSE.2011.37}

\bibitem[{{Vestergaard} \& {Wilkes}(2001)}]{Vestergaard2001}
{Vestergaard}, M., \& {Wilkes}, B.~J. 2001, \apjs, 134, 1, \dodoi{10.1086/320357}

\bibitem[{{Wang} {et~al.}(2019){Wang}, {Bai}, {Fan}, {Mao}, {Chang}, {Xin}, {Zhang}, {Lun}, {Wang}, {Zhang}, {Ying}, {Lu}, {Wang}, {Ji}, {Xiong}, {Yu}, {Ding}, {Ye}, {Xing}, {Yi}, {Xu}, {Zheng}, {Feng}, {He}, {Wang}, {Liu}, {Chen}, {Xu}, {Qin}, {Zhang}, {Tan}, {Li}, {Lou}, {Li}, \& {Liu}}]{Wang2019-LJ}
{Wang}, C.-J., {Bai}, J.-M., {Fan}, Y.-F., {et~al.} 2019, Research in Astronomy and Astrophysics, 19, 149, \dodoi{10.1088/1674-4527/19/10/149}

\bibitem[{{Wright} {et~al.}(2010){Wright}, {Eisenhardt}, {Mainzer}, {Ressler}, {Cutri}, {Jarrett}, {Kirkpatrick}, {Padgett}, {McMillan}, {Skrutskie}, {Stanford}, {Cohen}, {Walker}, {Mather}, {Leisawitz}, {Gautier}, {McLean}, {Benford}, {Lonsdale}, {Blain}, {Mendez}, {Irace}, {Duval}, {Liu}, {Royer}, {Heinrichsen}, {Howard}, {Shannon}, {Kendall}, {Walsh}, {Larsen}, {Cardon}, {Schick}, {Schwalm}, {Abid}, {Fabinsky}, {Naes}, \& {Tsai}}]{Wright2010}
{Wright}, E.~L., {Eisenhardt}, P. R.~M., {Mainzer}, A.~K., {et~al.} 2010, \aj, 140, 1868, \dodoi{10.1088/0004-6256/140/6/1868}

\bibitem[{{Yu} {et~al.}(2020){Yu}, {Martini}, {Davis}, {Gruendl}, {Hoormann}, {Kochanek}, {Lidman}, {Mudd}, {Peterson}, {Wester}, {Allam}, {Annis}, {Asorey}, {Avila}, {Banerji}, {Bertin}, {Brooks}, {Buckley-Geer}, {Calcino}, {Rosell}, {Carollo}, {Kind}, {Carretero}, {Cunha}, {D'Andrea}, {Costa}, {De Vicente}, {Desai}, {Diehl}, {Doel}, {Eifler}, {Flaugher}, {Fosalba}, {Frieman}, {Garc{\'\i}a-Bellido}, {Gaztanaga}, {Glazebrook}, {Gruen}, {Gschwend}, {Gutierrez}, {Hartley}, {Hinton}, {Hollowood}, {Honscheid}, {Hoyle}, {James}, {Kim}, {Krause}, {Kuehn}, {Kuropatkin}, {Lewis}, {Lima}, {Macaulay}, {Maia}, {Marshall}, {Menanteau}, {Miquel}, {M{\"o}ller}, {Plazas}, {Romer}, {Sanchez}, {Scarpine}, {Schubnell}, {Serrano}, {Smith}, {Smith}, {Soares-Santos}, {Sobreira}, {Suchyta}, {Swann}, {Swanson}, {Tarle}, {Tucker}, {Tucker}, \& {Vikram}}]{Yu2020}
{Yu}, Z., {Martini}, P., {Davis}, T.~M., {et~al.} 2020, \apjs, 246, 16, \dodoi{10.3847/1538-4365/ab5e7a}

\bibitem[{{Yuan} {et~al.}(2007){Yuan}, {Zdziarski}, {Xue}, \& {Wu}}]{Yuan2007}
{Yuan}, F., {Zdziarski}, A.~A., {Xue}, Y., \& {Wu}, X.-B. 2007, \apj, 659, 541, \dodoi{10.1086/512078}

\bibitem[{{Zu} {et~al.}(2013){Zu}, {Kochanek}, {Koz{\l}owski}, \& {Udalski}}]{Zu2013}
{Zu}, Y., {Kochanek}, C.~S., {Koz{\l}owski}, S., \& {Udalski}, A. 2013, \apj, 765, 106, \dodoi{10.1088/0004-637X/765/2/106}

\bibitem[{{Zu} {et~al.}(2011){Zu}, {Kochanek}, \& {Peterson}}]{Zu2011}
{Zu}, Y., {Kochanek}, C.~S., \& {Peterson}, B.~M. 2011, \apj, 735, 80, \dodoi{10.1088/0004-637X/735/2/80}

\end{thebibliography}
\bibliographystyle{aasjournal}

\end{document}